\documentclass[a4paper,11pt]{JHEP3}
\usepackage{amsmath,latexsym,amssymb}

\usepackage{graphicx}

\usepackage[latin1]{inputenc}
\usepackage{Macros}
\usepackage{nicefrac}
\usepackage{braket}
\usepackage{multirow}
\usepackage{arydshln} 

\usepackage{comment}

\title{Three-dimensional black holes from deformed anti~de~Sitter\thanks{
    Research partially supported by the EEC under the contracts
    \textsc{mext-ct}-2003-509661, \textsc{mrtn-ct}-2004-005104 and
    \textsc{mrtn-ct}-2004-503369.} }

\author{Stéphane~Detournay\footnote{``Chercheur FRIA", Belgium.}${\ }^\Diamond$,
  Domenico~Orlando${}^\spadesuit$,
  P.~Marios~Petropoulos${}^\spadesuit$, Philippe~Spindel${}^\Diamond$\\

\begin{itemize}
  
  \item  Mécanique et Gravitation, Université de Mons-Hainaut \\
  20 Place du Parc, 7000 Mons, Belgique
    
  \item   Centre de Physique Théorique, École Polytechnique\footnote{Unité
    mixte du CNRS et de l'École Polytechnique, UMR 7644.} \\
  91128 Palaiseau, France
\end{itemize}

\bigskip

E-mail: \email{Stephane.Detournay@umh.ac.be},
\email{orlando@cpht.polytechnique.fr},
\email{marios@cpht.polytechnique.fr}, \email{spindel@umh.ac.be} }

\abstract{We present new exact three-dimensional black-string backgrounds,
  which contain both NS--NS and electromagnetic fields, and generalize the
  \textsc{btz} black holes and the black string studied by Horne and
  Horowitz. They are obtained as deformations of the $SL(2,\setR)$ \textsc{wzw}
  model. Black holes resulting from purely continuous deformations possess
  true curvature singularities. When discrete identifications are
  introduced, extra chronological singularities appear, which under certain
  circumstances turn out to be naked. The backgrounds at hand appear in the
  moduli space of the $SL (2, \setR)$ \textsc{wzw} model. Hence, they provide
  exact string backgrounds and allow for a more algebraical (\textsc{cft})
  description. This makes possible the determination of the spectrum of
  primaries.}

\preprint{
  CPTH-RR022.0405 \\
  hep-th/0504231 \\}

\begin{document}

\setcounter{footnote}{0}
\renewcommand{\thefootnote}{\arabic{footnote}}
\setcounter{section}{0}


\newcommand{\p}{\ensuremath{\partial}}
\newcommand{\pb}{\ensuremath{\bar \partial}}
\newcommand{\SL}{\ensuremath{SL \left( 2, \setR \right)}}
\newcommand{\h}{\textsc{h}}
\section{Introduction}

The search for exact string backgrounds has been pursued over the past years
from various perspectives. Those investigations are motivated by
phenomenology, background-geometry analysis or, more recently, for
understanding holography beyond the usual supergravity approximation.

Anti-de Sitter backgrounds have played an important role in many respects.
Together with the spheres, they are the only maximally symmetric spaces
appearing naturally in string theory. They arise as near-horizon geometries
of distinguished brane configurations and offer the appropriate set up for
studying little-string theory, black-hole physics, \dots

The realization of anti-de Sitter spaces or spheres as string
backgrounds requires non-vanishing fluxes, which account for the
cosmological constant term in the low-energy equations of motion.
In general, those fluxes are of the Ramond--Ramond type, hence no
two-dimensional sigma-model is available. This happens indeed for
AdS$_5× S^5$ in type IIB or AdS$_4 × S^7$ in M-theory.  For
AdS$_3× S^3 × T^4$ (type IIA, B or heterotic), however, we have
the option to switch on a Neveu--Schwarz antisymmetric tensor
only. In this framework, the AdS$_3× S^3$ is the target space of
the $SL(2,\mathbb{R})× SU(2)$ Wess--Zumino--Witten model. The
latter has been studied
extensively~\cite{Antoniadis:1990mn,Petropoulos:1990fc,Boonstra:1998yu,Maldacena:1998bw,Giveon:1998ns,Israel:2003ry}.

Three-dimensional anti-de Sitter space provides a good laboratory for
studying many aspects of gravity and strings, including black-hole physics.
Locally anti-de Sitter three-dimensional black holes are obtained by
performing identifications in the original AdS$_3$ under discrete isometry
subgroups~\cite{Banados:1992wn,Banados:1993gq,Bieliavsky:2002ki,Bieliavsky:2003de}.
Those black holes (\textsc{btz}) have mass and angular momentum.
Generically, two horizons (inner and outer) mask the singularity, which
turns out to be a chronological singularity rather than a genuine curvature
singularity.

The two-dimensional sigma-model description of the AdS$_3$ plus
Kalb--Ramond field background allows for exact conformal
deformations, driven by integrable marginal
operators~\cite{Chaudhuri:1989qb,Hassan:1992gi,Giveon:1994ph,Forste:2003km,Israel:2003ry,Israel:2003cx,Israel:2004vv,Israel:2004cd}.
In general, a subgroup of the original isometry group survives
along those lines. Identification under discrete isometries is
thus legitimate and provides a tool for investigating new and
potentially interesting ``deformed \textsc{btz}'' geometries. The
latter may or may not be viable black holes, whereas black holes
may also appear by just deforming AdS$_3$ without further
surgery~\cite{Horne:1991gn}.

The aim of the present work is to clarify those issues, and reach
a global point of view on the geometries that emerge from the
$SL(2,\mathbb{R})$ WZW model, by using the above techniques.  This
will allow us to introduce new three-dimensional black hole
backgrounds that in general involve the presence of an electric
field. For these theories we give a complete \textsc{cft}
description, including an explicit expression for the spectrum of
string primaries. In particular, the usual black string
background~\cite{Horne:1991gn} will appear in this terms as a
special vanishing-field limit.  Carrying on identifications
\emph{à la} \textsc{btz} on these geometries will let us obtain
more black string and/or black hole backgrounds, generalizing the
one in~\cite{Banados:1992wn} and in~\cite{Horne:1991gn}, for which
we again provide a \textsc{cft} description. Not all the
backgrounds could be adapted to support the discrete
identifications. This will be stated in terms of a consistency
condition that has to be satisfied in order to avoid the presence
of naked (causal) singularities.

We will start with a quick overview of various distinct methods
based on Wess--Zumino--Witten models and aiming at generating new
exact \textsc{cft}s, that turn out to be equivalent to each other.
We will in particular exhibit their effect on the
$SL(2,\mathbb{R})$ WZW model. These results enable us to recast in
Sec. \ref{sec:blackstring} the three-dimensional black-string
solution of~\cite{Horne:1991gn}, as a patchwork of marginal
deformations of the $SL(2,\mathbb{R})$ WZW model. We clarify in
this way the role of the mass and charge parameters of the black
string.

Section~\ref{sec:two-parameter} is devoted to a two-parameter
deformation of $SL(2,\mathbb{R})$. This leads to a new family of
black strings, with NS--NS and electric field. We study the causal
structure of these black holes as well as their various charges.
They exhibit genuine curvature singularity hidden behind horizons.
In Sec.~\ref{sec:btz} we proceed with discrete identifications as
a solution-generating procedure
applied to the deformed AdS$_3$ -- wherever it is allowed by
residual symmetries.

After having stated the consistency conditions to be fulfilled in
order to avoid naked singularities, we find that time-like
chronological singularities protected by two horizons are
possible, while light-like singularities with only one horizon
appear as a limiting case.
Finally, in Sec.~\ref{sec:conf-field-theory} we determine the
spectrum of primaries, using standard \textsc{cft} techniques.



\section{Deformed \textsc{wzw} models: various perspectives}
\label{sec:wzw-deformations}

The power of \textsc{wzw} models resides in the symmetries of the theory.
Those impose strong constraints which allow quantum integrability
as well as a faithful description in terms of space--time fields
whose renormalization properties (at every order in
$\alpha^\prime$) are easily kept under
control~\cite{Knizhnik:1984nr,Leutwyler:1991tv,Tseytlin:1992ri}.

It is hence interesting to study the moduli space for these
models, aiming at finding less symmetric (and more interesting)
structures, that will hopefully enjoy analogous integrability and
space--time properties.

\subsection{Algebraic structure of current-current deformations}

In this spirit one can consider marginal deformations of the \textsc{wzw}
models obtained in terms of $\left(1,1\right)$ operators built as bilinears
in the currents:
\begin{equation}
\label{eq:Current-current-deform}
  \mathcal{O}(z, \bar z)= \sum_{ij} c_{ij} J^i \left( z \right) \tilde J^j \left( \bar z \right),
\end{equation}
where $J^i \left( z \right)$ and $\tilde J^j \left( \bar z\right)$
are respectively left- and right-moving currents. It is known that
this operator represents a truly marginal deformation if the
parameter matrix $c_{ij}$ satisfies appropriate
constraints~\cite{Chaudhuri:1989qb},
which are automatically satisfied for any value of $c_{ij}$,
whatever the algebra,
 if $J^i$ and $\tilde J^j$ live on a torus. Hence, we get as
moduli space continuous surfaces of exact models\footnote{Although
for special values of the level $k$
  the theory contains other operators with the right conformal weights, it is
  believed that only current-current operators give rise to truly marginal
  deformations, \emph{i.e.} operators that remain marginal for finite values
  of the deformation parameter.}.

From the \textsc{cft} point of view, it is
known~\cite{Forste:2003km} that the effect of the deformation is
completely captured by an $O (d,\bar d) $ pseudo-orthogonal
transformation of the charge lattice $\Lambda \subset
\mathfrak{h}^\ast × \bar {\mathfrak{h}}^\ast$ of the abelian
sector of the theory ($\mathfrak{h} \subset \mathfrak{g}$ and
$\bar{\mathfrak{h}} \subset \bar {\mathfrak{g}}$ being abelian
subalgebras of the undeformed \textsc{wzw} model $\mathfrak{g} ×
\bar{\mathfrak{g}}$ algebra). Moreover, since the charges only
characterize the $\mathfrak{h} × \bar{\mathfrak{h}}$ modules up to
automorphisms of the algebras, $O (d) × O (\bar d) $
transformations don't change the \textsc{cft}. Hence the
deformation space is given by:
\begin{equation}
  D_{\mathfrak{h}, \bar{\mathfrak{h}}} \sim O (d, \bar d) / \left(O(d)
  × O(\bar d) \right).
\end{equation}
The moduli space is obtained out of $D_{\mathfrak{h},
\bar{\mathfrak{h}}}$ after the identification of the points giving
equivalent \textsc{cft}s\footnote{Although we will concentrate on
\textsc{wzw} models it is worth to emphasize that this
construction is more general.}.

In the case of \textsc{wzw} models on compact groups, all maximal abelian
subgroups are pairwise conjugated by inner automorphisms. This implies that the
complete deformation space is $D = O (d, d) / \left(O(d) × O(d) \right)$
where $d$ is the rank of the group. The story is different for
non-semisimple algebras, whose moduli space is larger, since we get different $O
(d, \bar d) / \left( O(d) × O(\bar d) \right)$
deformation spaces for each (inequivalent) choice of the
abelian subalgebras $\mathfrak{h} \subset \mathfrak{g}$ and $\bar{\mathfrak{h}}
\subset \bar{\mathfrak{g}}$.

An alternative way of describing current-current deformations
comes from the so-called parafermion decomposition. The
highest-weight representation for a $\hat{\mathfrak{g}}_k$ graded
algebra can be decomposed into highest-weight modules of a Cartan
subalgebra $\hat{\mathfrak{h}} \subset \hat{\mathfrak{g}}_k$ as
follows~\cite{Gepner:1986hr,Gepner:1987sm}:
\begin{equation}
  \mathcal{V}_{\hat \lambda } \simeq \bigoplus_{\mu \in \Gamma_k}
\mathcal{V}_{\hat \lambda, \mu} \otimes \bigoplus_{\delta \in Q_l
(\mathfrak{g})} \mathcal{V}_{\mu + k \delta},
\end{equation}
where $\hat \lambda$ is an integrable weight of
$\hat{\mathfrak{g}}_k$, $\mathcal{V}_{\hat \lambda, \mu}$ is the
highest-weight module for the generalized $\hat{\mathfrak{g}}_k /
\hat{\mathfrak{h}}$ parafermion, $Q_l ( \mathfrak{g} )$ is the
long-root lattice and $\Gamma_k = P ( \mathfrak{g} )/Q_l(
\mathfrak{g} )$ with $P ( \mathfrak{g} ) $ the weight lattice. As
a consequence, the \textsc{wzw} model based on
$\hat{\mathfrak{g}}_k$ can be represented as an orbifold model:
\begin{equation}
  \label{eq:ParafermionRep}
  \hat{\mathfrak{g}}_k \simeq \left( \hat{\mathfrak{g}}_k/ \hat{\mathfrak{h}}
\otimes t_{\Lambda_k} \right) / \Gamma_k,
\end{equation}
where $t_{\Lambda_k}$ is a toroidal \textsc{cft} with charge
lattice, included in the $\hat{\mathfrak{g}}_k$ one, defined as
$\Lambda_k = \set{ \left( \mu, \bar \mu \right) \in P (
\hat{\mathfrak{g}}) × P (\hat{\mathfrak{g}} ) | \mu - \bar
\mu = k Q_l (\hat{\mathfrak{g}})}$. The advantage given by using
this representation relies on the fact that $\Gamma_k$ acts
trivially on the coset and toroidal model algebras; then, if we
identify $\hat{\mathfrak{h}} $ and $\bar{\hat{\mathfrak{h}}}$ with
the graded algebras of $t_{\Lambda_k}$, the deformation only acts
on the toroidal lattice and the deformed model can again be
represented as an orbifold:
\begin{equation}
  \label{eq:DefParafermionRep}
  \hat{\mathfrak{g}}_k ( \mathcal{O} ) \simeq \left( \hat{\mathfrak{g}}_k/
\hat{\mathfrak{h}} \otimes t_{\mathcal{O}\Lambda_k} \right) /
\Gamma_k,
\end{equation}
where $\mathcal{O}$ is an operator in the moduli space. This representation
is specially useful because it allows to easily single out the sector of the
theory that is affected by the deformation. As we'll see in the next section
this simplifies the task of writing the corresponding Lagrangian.



In the following we will separate (somehow arbitrarily) this kind of
deformations into two categories:
those who give rise to \emph{symmetric} deformations, \emph{i.e.} the ones where
$c_{ij} = \delta_{ij}$ and $J^i \left( z \right)$ and $\tilde J^j \left(
  \bar z \right)$ represent the same current in the two chiral sectors of
the theory and the \emph{asymmetric} ones where the currents are
different and in general correspond to different subalgebras. In
some ways this distinction is arbitrary, since both symmetric and
asymmetric deformations act as $O \left( d, \bar d \right)$
rotations on the background fields. It is nonetheless interesting
to single out the asymmetric case. In the particular situation,
when one of the two currents belongs to an internal $U \left( 1
\right)$ (coming from the gauge sector in the heterotic or simply
from any $U\left( 1 \right)$ subalgebra in the type II), it is
particularly simple to study the effect of the deformation, even
from the space--time field point of view; there, the expressions
for the background fields are exact (at all order in
$\alpha^\prime$ and for every value of the level
$k$)~\cite{Israel:2004vv}.

\subsection{Background fields and symmetric deformations}
\label{sec:backgr-fields-symm}

\subsubsection*{General Construction}

Symmetric deformations (also called \emph{gravitational}) are
those that have received by far the most attention in literature.
Specializing Eq.~\eqref{eq:Current-current-deform} to the case of
one only current we can write the small deformation Lagrangian as:
\begin{equation}
  S = S_{\textsc{wzw}} + \delta \kappa^2 \int \di^2 z \: J (z ) \bar J ( \bar z )
\end{equation}
This infinitesimal deformation has to be integrated in order to give a
Lagrangian interpretation to the \textsc{cft} described in the previous
section. Different approaches are possible, exploiting the different
possible representations described above.
\begin{itemize}
\item A possible way consists in implementing an $O(d,d)$ rotation on the
background fields~\cite{Hassan:1992gi}. More precisely, one has to
identify a coordinate system in which the background fields are
independent of $d$ space dimensions and metric and $B$ field are
written in a block diagonal form. In this way the following matrix
is defined:
\begin{equation}
  M = \left( \begin{tabular}{c|c}
      $\hat{g}^{-1}$ &  $-\hat{g}^{-1} \hat{B}$ \\ \hline
      $\hat{B} \hat{g}^{-1}$ & $\hat{g} - \hat{B}\hat{g}^{-1}\hat{B}$
    \end{tabular}
  \right),
\end{equation}
where $\hat g $ and $\hat B$ are the pull-backs of the metric and
Kalb--Ramond field on the $p$ selected directions. Then the action
of the $O(d,d)$ group on these fields and dilaton is given by:
\begin{align}
  M & \to M^\prime = \Omega M \Omega^{t}, \label{eq:OddMetric}\\
  \Phi & \to \Phi^\prime = \Phi - \frac{1}{4} \log \left( \frac{\det \hat
g}{\det \hat g^\prime} \right),\label{eq:OddDilaton}
\end{align}
where $\hat g^\prime$ is the metric after the
transformation~\eqref{eq:OddMetric} and $\Omega \in O (d,d)$. It
must be emphasized that this transformation rules are valid at the
lowest order in $\alpha^\prime$ (but at all orders in the
deformation parameters). So, although the model is exact, as we
learn from the \textsc{cft} side, the field expressions that we
find only are true at leading order in $\alpha^\prime$.
\item An alternative approach uses the parafermion representation
  Eq.~\eqref{eq:DefParafermionRep} (see \emph{e.g.}~\cite{Forste:2003km}). In
  practice this amounts to writing an action as the sum of the $G/H$ parafermion
  and a deformed $H$ part and finding the appropriate T-duality transformation
  (realizing the orbifold) such that for zero deformation the \textsc{wzw} on $G$
  is recovered, in accordance with Eq.~\eqref{eq:ParafermionRep}.
\item Finally, another point of view (inspired by the parafermionic
  representation), consists in identifying the deformed model with a $\left(
    G × H \right) / H $ coset model, in which the embedding of the dividing
  group has a component in both factors~\cite{Giveon:1994ph}. The gauging of
  the component in $G$ gives the parafermionic sector, the gauging of the
  component in $H$ gives the deformed toroidal sector and the coupling term
  (originating from the quadratic structure in the fields introduced for the
  gauging) corresponds to the orbifold projection\footnote{An
    instanton-correction-aware technique that should overcome the first
    order in $\alpha^\prime$ limitation for gauged models has been proposed
    in~\cite{Tseytlin:1994my}. In principle this can be used to get an
    all-order exact background when we write the deformation as a gauged
    model. We will not expand further in this direction, that could
    nevertheless be useful to address issues such as the stability of the
    black string (see Sec. \ref{sec:blackstring}).}.
\end{itemize}

\subsubsection*{The $\SL$ case}

In the present  work, we want to concentrate on the deformations
of $SL (2, \setR )$. Symmetric deformations of this \textsc{wzw} model are
known in the literature. The group manifold of $SL(2,\mathbb{R})$
is anti de Sitter in three dimensions. Metric and antisymmetric
tensor read (in Euler coordinates, see App. \ref{antidss}):
\begin{subequations}
  \begin{align}
    \di s^2&= L^2\left[\di \rho^2 +
    \sinh^2 \rho \, \di \phi^2 - \cosh ^2 \rho  \, \di \tau ^2 \right],     \\
    H_{[3]} &= L^2\sinh 2\rho
      \di \rho \land \di \phi  \land \di \tau,
  \end{align}
\end{subequations}
with $L$ related to the level of $SL(2,\mathbb{R})_k$ as usual:
$L=\sqrt{k+2}$. In the case at hand, three different
lines of symmetric deformations arise due to the presence of
time-like ($J^3$, $\bar J^3$), space-like ($J^1$, $\bar J^1$,
$J^2$, $\bar J^2$), or null generators
\cite{Forste:2003km,Forste:1994wp,Israel:2003ry}. The residual
isometry is $U(1) × U(1)$ that can be time-like $(L_3, R_3 )$,
space-like $(L_2, R_2 )$ or null $(L_1 + L_3, R_1 + R_3 )$
depending on the deformation under consideration.


The \emph{elliptic deformation} is driven by the $J^3\bar J^3$
bilinear. At first order in $\alpha^{\prime}$ the background
fields are given by\footnote{The extra index ``3" in the
  deformation parameter $\kappa$ reminds that the deformation refers here to $J^3
  \bar J^3$.}:
\begin{subequations}
  \begin{align}
    \di s^2&= k \left[\di \rho^2 + \frac{\sinh^2 \rho \, \di \phi^2 -\kappa_3^2 \cosh
        ^2 \rho \, \di \tau ^2
      }{\Theta_{\kappa_3} (\rho )} \right],\label{eq:J3J3-metric}\\
    H_{[3]} & = k \frac{\kappa_3^2\sinh 2\rho}{\Theta_{\kappa_3} (\rho )^2 }
    \di \rho \land \di \phi  \land \di \tau,\\
    { \mathrm e}^{-2 \Phi} &= \frac{\Theta_{\kappa_3} (\rho )}{\kappa_3}.
  \end{align}
\end{subequations}
where $\Theta_{\kappa_3} (\rho ) = \cosh ^2 \rho -\kappa_3^2
\sinh^2 \rho$ and, of course, $\Phi$ is defined up to an additive
constant. At extreme deformation ($\kappa_3^2 \to 0$), a time-like
direction decouples and we are left with the axial\footnote{The
deformation parameter has two T-dual branches.
  The extreme values of deformation correspond to the axial or vector
  gaugings. The vector gauging leads to the \emph{trumpet}.  For the
  $SU(2)_k / U(1)$, both gaugings correspond to the \emph{bell}.}
${SL(2,\mathbb{R})_k / U(1)_{\text{time}}}$. The target space of the latter
is the \emph{cigar} geometry (also called Euclidean two-dimensional black
hole):
\begin{eqnarray}
{\mathrm e}^{-2 \Phi}&\sim & \cosh^2 \rho,\\
\di s^2&=&k \left[ \di \rho^2+\tanh^2\rho \, \di \phi ^2 \right],
\end{eqnarray}
($0\leq \rho < \infty$ and $0\leq \phi \leq 2\pi$).


Similarly, with $J^2\bar J^2$ one generates the \emph{hyperbolic
deformation}. This allows to reach the Lorentzian two-dimensional
black hole times a free space-like line. Using the coordinates
defined in Eq.~(\ref{sphan}), we find:
\begin{subequations}
\label{eq:J2J2-deform}
  \begin{align}
    \di s^2&= k \left[- \di \beta^2 + \frac{\sin^2 \beta \, \di \varphi^2
    + \kappa_2^2 \cos ^2 \beta  \, \di \psi^2}{\Delta_{\kappa_2} (\beta)}\right], \label{eq:J2J2-metric}\\
    H_{[3]} & = k \frac{\kappa_2^2\sin 2\beta}{\Delta_{\kappa_2} (\beta)^2 }
      \di \beta \land \di \psi  \land \di \phi,\\
  { \mathrm e}^{-2\Phi} &= \frac{\Delta_{\kappa_2} (\beta)}{\kappa_2},
  \end{align}
\end{subequations}
where $\Delta_{\kappa_2} (\beta) = \cos ^2 \beta + \kappa_2^2
\sin^2 \beta$.  This coordinate patch does not cover the full
$\mathrm{AdS}_3$. We will expand on this line in
Sec.~\ref{sec:blackstring}.


Finally, the bilinear $\left(J^1 + J^3\right)\left(\bar J^1 + \bar
  J^3\right)$ generates the \emph{parabolic deformation}. Using Poincaré
coordinates
(Eqs.~(\ref{eq:ads-poinc-tr})--(\ref{eq:ads-poinc-volume}))\footnote{Note
  that $x^± = X ± T$.} we obtain:
\begin{subequations}
\label{eq:null-deform}
  \begin{align}
    \di s^2&= k \left[\frac{\di u^2}{u^2}+
    \frac{\di X ^2- \di T^2}{u^2 + 1/\nu}
    \right], \label{eq:null-metric}\\
    H_{[3]} & = k \frac{2u}{\left(u^2 + 1/ \nu \right)^2}
      \di u \land \di T  \land \di X,\\
  { \mathrm e}^{-2\Phi} &= \frac{u^2 + 1/\nu }{u^2}.
  \end{align}
\end{subequations}
The deformation parameter is $1/\nu$. At infinite value of the
parameter $\nu$, we recover pure AdS$_3$; for $\nu \to 0$, a whole
light-cone decouples and we are left with a single direction and a
dilaton field, linear in this direction.

The physical interpretation of the parabolic deformation is far
reaching, when AdS$_3$ is considered in the framework of the
\textsc{ns5/f1} near-horizon background, AdS$_3× S^3 × T^4$. In
this physical set-up, the parameter $\nu$ is the density of
\textsc{f1}'s (number of fundamental strings over the volume of
the four-torus $T^4$)~\cite{Israel:2003ry,Kiritsis:2003cx}.
At infinite density, the background is
indeed AdS$_3× S^3 × T^4$. At null density, the geometry becomes
$\mathbb{R}^{1,2}× S^3 × T^4$ plus a linear dilaton and a
three-form on the $S^3$.

\subsection{Background fields and asymmetric deformations}
\label{sec:backgr-fields-asymm}

\subsubsection*{General construction}

Consider the case of $G = G^\prime × U \left( 1 \right)^r $,
$H = U \left( 1 \right)^r $ where $r = \rank \left( G \right)$
embedded such as $\epsilon_{\textsc{l}} \left( H \right) \subset
G^\prime $ and $\epsilon_{\textsc{r}} \left( H \right) = G^{\prime
\prime} = U \left( 1 \right)^r$. To clarify the notation we can
write the deformation operator as:
\begin{equation}
  \mathcal{O}(z, \bar z)  = \sum_{a=1}^r \h_a J^a \left( z \right) \db X^a
\end{equation}
where $X^a \left( z, \bar z\right)$ results from the bosonisation
of the right current.  Using \emph{e.g.} Kaluza-Klein
reduction~\cite{Horowitz:1995rf,Kiritsis:1994ta,Israel:2004cd},

 one shows that the
effect of the deformation on the background fields (identified as those
living in the $G^\prime $ sector) is the following:
\begin{subequations}
  \label{eq:asym-deform}
  \begin{align}
    G_{\mu \nu } &= \mathring G_{\mu \nu } - 2 \sum_{i=a}^r \h_a^2 J^a_{\mu }
    J^a_{\nu} \label{eq:asym-metric},\\
    B_{\mu \nu } &= \mathring B_{\mu \nu}, \\
    A^a_{\mu } &= \h_a \sqrt{\frac{2k}{k_g}} J_\mu^a,
  \end{align}
\end{subequations}
where $\mathring G_{\mu \nu}$ and $\mathring B_{\mu \nu }$ are the initial,
unperturbed background fields that are expressed in terms of the $g \in G^\prime
$ group element as follows:
\begin{subequations}
  \begin{align}
    \mathring G_{\mu \nu } \di x^\mu \di x^\nu &= \braket{ g^{-1} \di g , g^{-1}
      \di g}, \\
    \mathring B_{\mu \nu } \di x^\mu \land x^\nu &= g^{-1} \di g \land g^{-1} \di g.
  \end{align}
\end{subequations}
No dilaton is present (as a consequence of the fact that the Ricci scalar
for these deformed systems remains constant) and these semiclassical
solutions can be promoted to exact ones just by remarking that the effect of
the renormalisation simply boils down to the shift $k \to k + c_{G^\prime}$ where
$c_{G^\prime}$ is the dual Coxeter number, just as in the case of the
unperturbed \textsc{wzw} model.

\subsubsection*{The $\SL$ case}

We now apply the above to the $SL(2,\mathbb{R})$ case. As previously, three
asymmetric deformations are available: the elliptic, the
hyperbolic and the parabolic.

The \emph{elliptic deformation} is generated by a bilinear where
the left current is an $SL(2,\mathbb{R})_k$ time-like current. The
background field is magnetic and the residual symmetry is
$U(1)_{\text{time}} × SL(2,\mathbb{R})$ generated by $\{L_3,
R_1, R_2, R_3 \}$ (see App. \ref{antidss}). The metric reads (in
elliptic coordinates):
\begin{equation}
  \di s^2= \frac{k}{4} \left[ \di \rho^2 + \cosh^2 \rho  \di \phi^2 -
    \left( 1 + 2 \h^2\right) \left( \di t + \sinh \rho \di
      \phi \right)^2 \right],
  \label{dsnecogo}
\end{equation}
where $\partial_t$ is the Killing vector associated with the
$U(1)_{\text{time}}$. This AdS$_3$ deformation was studied in
\cite{Rooman:1998xf} as a \emph{squashed anti de Sitter} and in
\cite{Israel:2003cx,Israel:2004vv} from the string theory point of view. It has
curvature
\begin{equation}
  \mathcal{R}=-\frac{2}{k}(3 - 2\h^2). \label{curnecogo}
\end{equation}
Here, it comes as an \emph{exact string solution} (provided $k\to
k +2$) together with an NS three-form and a magnetic field:
\begin{subequations}
  \begin{align}
  H_{[3]} &= \di B - \frac{k_g}{4} A \land \di A =
  - \frac{k}{4}\left( 1+ 2\h^2\right) \cosh \rho \di \rho \land \di \phi \land \di
  t,  \label{adsmagH}
  \\
    A &= \h \sqrt{\frac{2k}{k_g}} \left(\di t + \sinh \rho \di \phi
  \right).
    \label{adsmag}
  \end{align}
\end{subequations}
For $\h^2>0$ (unitary region), the above metric is pathological because it has
topologically trivial closed time-like curves passing through any
point of the manifold. Actually, for $\h^2=1/2$ we recover exactly
the Gödel space, which is a well-known example of pathological
solution of Einstein--Maxwell equations.

The \emph{hyperbolic deformation} can be studied in a similar
fashion, where the left current in the bilinear is an
$SL(2,\mathbb{R})_k$ space-like current. In hyperbolic
coordinates:
\begin{equation}
  \di s^2= \frac{k}{4}\left[ \di r^2 - \cosh^2 r \di \tau^2 +
    \left( 1-2\h^2\right) \left( \di x + \sinh r \di \tau
    \right)^2\right],
  \label{dsnecoma}
\end{equation}
where $\partial_x$ generates a $U(1)_{\text{space}}$. The total
residual symmetry is $U(1)_{\text{space}} ×
SL(2,\mathbb{R})$, generated by $\{L_2, R_1, R_2, R_3 \}$, and
\begin{equation}
  \mathcal{R}=-\frac{2}{k}\left(3+2\h^2\right).\label{Rel}
\end{equation}
The complete string background now has an NS three-form and an
electric field:
\begin{subequations}
  \begin{align}
    H_{[3]} &= \frac{k}{4} \left(1-2\h^2\right) \cosh r \di r \land
      \di \tau \land \di  x,\label{adselH}\\
    A &= \h \sqrt{\frac{2k}{k_g}} \left( \di x + \sinh r \di \tau
  \right).\label{adsel}
  \end{align}
\end{subequations}

The background at hand is free of closed time-like curves. The
squashed AdS$_3$ is now obtained by going to the AdS$_3$ picture
as an $S^1$ fibration over an AdS$_2$ base, and modifying the
$S^1$ fiber. The magnitude of the electric field is limited at
$\h_{\text{max}}^2 = 1/2$, where it causes the degeneration of the
fiber, and we are left with an AdS$_2$ background with an electric
monopole; in other words, a geometric coset $SL(2,\mathbb{R})/
U(1)_{\text{space}}$.

The string spectrum of the above deformation is accessible by
conformal-field-theory methods. It is free of tachyons and a whole
tower of states decouples at the critical values of the electric
fields. Details are available in~\cite{Israel:2004vv}.


Finally, the \emph{parabolic deformation} is generated
by a null $SL(2,\mathbb{R})_k$ current times some internal
right-moving current. The deformed metric reads, in Poincar{é}
coordinates:
\begin{equation}
  \di s^2 =k\left[\frac{ \di u^2 }{u^2}+ \frac{\di x^+ \di x^-
    }{u^2}-2\h^2 \left(\frac{\di x^+}{u^2}\right)^2 \right],
  \label{dsemdef}
\end{equation}
and the curvature remains unaltered $\mathcal{R}=-6/k$. This is
not surprising since the resulting geometry is a plane-wave like
deformation of AdS$_3$. The residual symmetry is
$U(1)_{\text{null}} × SL(2,\mathbb{R})$, where the $U(1)_{\text{
null}}$ is generated by $\partial_- = -L_1-L_3$.

The parabolic deformation is somehow peculiar. Although it is continuous,
the deformation parameter can always be re-absorbed by a redefinition of the
coordinates\footnote{This statement holds as long as these coordinates are
  not compact.  After discrete identifications have been imposed (see Sec.
  \ref{sec:btzas})), $\h$ becomes a genuine continuous parameter.}: $x^+ \to
x^+ / \abs{\h}$ and $x^-\to x^- \abs{\h}$. Put differently, there
are only three truly different options: $\h^2 = 0,  1$. No
limiting geometry emerges in the case at hand.

As expected, the gravitational background is accompanied by an NS
three-form (unaltered) and an electromagnetic wave:
\begin{equation}
  A = 2 \sqrt{\frac{2k}{k_g}} \h \frac{\di x^+}{u^2}.\label{adsem}
\end{equation}

A final remark is in order here, which holds for all three asymmetric
deformations of $SL(2,\mathbb{R})$. The background electric or magnetic
fields that appear in these solutions (Eqs.~(\ref{adsmag}), (\ref{adsel})
and (\ref{adsem})) diverge at the boundary of the corresponding spaces.
Hence, these fields cannot be considered as originating from localized
charges.



\section{The three-dimensional black string revisited}
\label{sec:blackstring}

The AdS$_3$ moduli space contains black hole geometries. This has been known
since the most celebrated of them -- the two-dimensional $SL (2, \setR )/U(1)$
black hole -- was found by Witten~\cite{Witten:1991yr,Dijkgraaf:1992ba}.
Generalisations of these constructions to higher dimensions have been considered
in~\cite{Horne:1991gn,Gershon:1991qp,Horava:1991am,Klimcik:1994wp}. The
three-dimensional black
string~\cite{Horne:1991gn,Horne:1991cn,Horowitz:1993jc} has attracted much
attention, for it provides an alternative to the Schwarzschild black hole in
three-dimensional asymptotically flat geometries\footnote{Remember that the
  \emph{no hair} theorem doesn't hold in three
  dimensions~\cite{Israel:1967wq,Heusler:1998ua,Gibbons:2002av}.}. In this
section we want to show how this black string can be interpreted in terms of
marginal deformations of $SL ( 2, \setR)$, which will enable us to give an
expression for its string primary states (Sec.~\ref{sec:conf-field-theory}).

In \cite{Horne:1991gn} the black string was obtained as an $\left( SL (2,
  \setR) × \setR \right) /\setR$ gauged model. More precisely, expressing $g \in
SL(2,\setR) × \setR$ as:
\begin{equation}
  g = \begin{pmatrix}
    a & u & 0 \\
    -v & b & 0 \\
    0 & 0 & {\rm e}^x
  \end{pmatrix},
\end{equation}
the left and right embeddings of the $\setR$ subgroup are given by:
\begin{align}
  \begin{split}
    \epsilon_L : \setR &\to \SL × \setR \\
    \lambda &\mapsto \begin{pmatrix}
      {\rm e}^\lambda & 0 & 0 \\
      0 & {\rm e}^{-\lambda} & 0 \\
      0 & 0 & 1
    \end{pmatrix}
  \end{split}
  \begin{split}
    \epsilon_R : \setR &\to \SL × \setR \\
    \lambda &\mapsto \begin{pmatrix}
      1 & 0 & 0 \\
      0 & 1 & 0 \\
      0 & 0 & {\rm e}^\lambda
    \end{pmatrix}
  \end{split}
\end{align}
so that in $\epsilon_L$ one recognises the $\SL$ subgroup generated by the
$J^2$ current while $\epsilon_R$ describes an embedding in the $\setR$ part
alone. From the discussion in Sec.~\ref{sec:backgr-fields-symm}, we see that
performing this gauging is just one of the possible ways to recover the $J^2
\bar J^2$ symmetrically deformed $\SL$ geometry. More specifically, since
the gauged symmetry is axial ($g \to h g h$), it corresponds (in our
notation) to the $\kappa_2 < 1$ branch of the deformed geometry\footnote{The $R \gtrless 1$
  convention is not univocal in literature.} in Eq. \eqref{eq:J2J2-metric}.
One can find a coordinate transformation allowing to pass from the usual
black-string solution
\begin{subequations}
\label{eq:black-string}
  \begin{align}
    \di s^2 &= \frac{k}{4}\left[-\left(1-\frac{1}{r}\right) \di t^2 + \left(1-\frac{\mu^2}{
      r}\right) \di x^2 + \left(1-\frac{1}{r}\right)^{-1}
      \left(1-\frac{\mu^2}{r}\right)^{-1} \frac{\di r^2}{r^2}\right],\label{eq:black-string-met} \\
    H &= \frac{k}{4} \frac{\mu}{r} \di r \land \di x \land \di t, \\
    \mathrm{e}^{2 \Phi} &= \frac{\mu}{r}
  \end{align}
\end{subequations}
to our (local) coordinate system, Eq.~\eqref{eq:J2J2-deform}. The attentive
reader might now be puzzled by this equivalence between a one-parameter
model such as the symmetrically deformed model and a two-parameter one such
as the black string in its usual coordinates (in
Eqs.~\eqref{eq:black-string} we redefined the $r$ coordinate as $r\to r/M$
and then set $\mu = Q/M$ with respect to the conventions
in~\cite{Horne:1991gn}). A point that it is interesting to make here is that
although, out of physical considerations, the black string is usually
described in terms of two parameters (mass and charge), the only physically
distinguishable parameter is their ratio $\mu = Q/M$ that coincides with our
$\kappa_2$ parameter. In Sec.~\ref{sec:two-parameter} we will introduce a
different (double) deformation, this time giving rise to a black hole
geometry depending on two actual parameters (one of which being related to
an additional electric field).


As we remarked above, the axial gauging construction only applies
for $\mu <1 $, while, in order to obtain the other $\kappa_2 > 1$
branch of the $J^2 \bar J^2$ deformation, one should perform a
vector gauging. On the other hand, this operation, that would be
justified by a \textsc{cft} point of view, is not natural when one
takes a more geometrical point of view and writes the black string
metric as in Eq.~\eqref{eq:black-string-met}. In the latter, one
can study the signature of the metric as a function of $r$ in the
two regions $\mu^2 \gtrless 1 $, and find the physically sensible
regions (see Tab.~\ref{tab:bs-signature}).

\TABLE{ \begin{tabular}{|c|c|c|c|c|c|c|} \hline \multirow{2}{*}{$\mu$}&
    \multirow{2}{*}{name} & $\di t^2 $ & $\di x^2 $ & $\di r^2$ &
    \multirow{2}{*}{range} & \multirow{2}{2.5cm}{\textsc{cft}
      interpretation} \\
    \cline{3-5} && $ - \left( 1- \frac{1}{r} \right) $ & $1- \frac{\mu^2}{
      r}$ & $ \left( 1- \frac{1}{r} \right)^{-1} \left( 1- \frac{\mu^2}{r}
    \right)^{-1} $ & &\\ \hline \hline \multirow{3}{*}{$\mu^2 >1$}& $\left(
      c^+ \right)$ & $-$ & $+$ & $+$ &$ r
    > \mu^2$ & $J^3 \bar J^3$, $\kappa_3>1$  \\
    \cline{2-2}\cdashline{3-3} \cline{4-7} &$\left( b^+ \right)$ & $-$ & $-$
    & $-$ &$1< r< \mu^2$ &\\ \cline{2-3} \cdashline{4-4} \cline{5-7} &$\left(
      a^+ \right)$ & $+$ & $-$ & $+$ & $0< r<1$ & $J^3 \bar J^3$, $\kappa_3<1$
    \\ \hline \hline \multirow{3}{*}{$\mu^2 < 1$}& $\left( a^- \right)$ & $+$
    & $-$ & $+$ & $0 < r < \mu^2$ & \multirow{3}{*}{$J^2 \bar J^2$, $\kappa_2 <
      1$} \\ \cline{2-2}\cdashline{3-3} \cline{4-6} &$\left( b^- \right)$ &
    $+$ & $+$ & $-$ &$\mu^2 < r < 1 $ & \\ \cline{2-3} \cdashline{4-4}
    \cline{5-6} &$\left( c^- \right)$ & $-$ & $+$ & $+$ &$ r > 1$ & \\
    \hline
  \end{tabular}
  \caption{Signature for the black-string metric as a function of $r$, for
    $\mu^2 \gtrless 1 $.}
  \label{tab:bs-signature}}

Our observations are the following:
\begin{itemize}
\item The $\mu^2 < 1$ branch always has the correct $\left( -, +, + \right)
  $ signature for any value of $r$, with the two special values $r = 1 $ and
  $r = \mu^2 $ marking the presence of the horizons that hide the
  singularity in $r=0$.
\item The $\mu^2> 1$ branch is different. In particular we see that there
  are two regions: $\left(a^+\right)$ for $0<r<1$ and $\left( c^+ \right)$
  for $r > \mu^2$ where the signature is that of a physical space.
\end{itemize}
A fact deserves to be emphasized here: one should notice that while for
$\mu^2 < 1$ we obtain three different regions of the same space, for $\mu^2 >
1$ what we show in Tab.~\ref{tab:bs-signature} really are three different
spaces and the proposed ranges for $r$ are just an effect of the chosen
parameterization. The $\left( a^+ \right), \kappa_3 < 1 $ and $\left(
  c^+\right), \kappa_3 > 1 $ branches are different spaces and not different
regions of the same one and one can choose in which one to go when
continuing to $\mu > 1$.

But there is more. The $\mu^2 > 1 $ region is obtained via an
analytic continuation with respect to the other branch, and this
analytic continuation is precisely the one that interchanges the
roles of the $J^2$ and the $J^3$ currents. As a result, we pass
from the $J^2 \bar J^2$ line to the $J^3 \bar J^3$ line. More
precisely the $\left(c^+\right)$ region describes the ``singular"
$\kappa_3 > 1 $ branch of the $J^3 \bar J^3 $ deformation
(\textit{i.e.} the branch that includes the $r=0$ singularity) and
the $\left(a^+\right)$ region describes the regular $\kappa_3<1$
branch that has the \emph{cigar} geometry as $\kappa_3 \to 0$
limit. Also notice that the regions $r<0$ have to be excluded in
order to avoid naked singularities (of the type encountered in the
Schwarzschild black hole with negative mass). The black string
described in~\cite{Horne:1991gn} covers the regions
$\left(a^-\right), \left(b^-\right), \left(c^-\right),
\left(a^+\right)$.

Our last point concerns the expectation of the genuine $\mathrm{AdS}_3$
geometry as a zero-deformation limit of the black-string metric, since the
latter turns out to be a marginal deformation of AdS$_3$ with parameter
$\mu$. The straightforward approach consists in taking the line element in
Eq.~\eqref{eq:black-string-met} for $\mu = 1$. It is then puzzling that the
resulting extremal black-string geometry \emph{is not} $\mathrm{AdS}_3 $.
This apparent paradox is solved by carefully looking at the coordinate
transformations that relate the black-string coordinates $(r,x,t)$ to either
the Euler coordinates $(\rho, \phi, \tau)$ (\ref{euler}) for the $J^3 \bar
J^3$ line, or the hyperbolic coordinates $(y,x,t)$ (\ref{sphan}) for the
$J^2 \bar J^2$ line. These transformations are singular at $\mu = 1$, which
therefore corresponds neither to $\kappa_3 =1$ nor to $\kappa_2 =1$.  Put differently,
$\mu = 1$ is not part of a continuous line of deformed models but marks a
jump from the $J^2 \bar J^2 $ to the $J^3 \bar J^3 $ lines.

The extremal black-string solution is even more peculiar.
Comparing Eqs. (\ref{eq:black-string}) at $\mu = 1$ to Eqs.
(\ref{eq:null-deform}), which describe the symmetrically
null-deformed $SL(2,\mathbb{R})$, we observe that the two
backgrounds at hand are related by a coordinate transformation,
provided $\nu = -1$.

The black string background is therefore entirely described in
terms of $SL(2,\mathbb{R})$ marginal symmetric deformations, and
involves all three of them. The null deformation appears, however,
for the extremal black string only and at a negative value of the
parameter $\nu$. The latter is the density of fundamental strings,
when the deformed AdS$_3$ is considered within the \textsc{ns5/f1}
system. This might be one more sign pointing towards a possible
instability in the black string \cite{Gregory:1993vy}.

Notice finally that expressions~\eqref{eq:black-string} receive
$1/k$ corrections. Those have been computed
in~\cite{Sfetsos:1992yi}. Once taken into account, they contribute
in making the geometry smoother, as usual in string theory.

\section{The two-parameter deformations}
\label{sec:two-parameter}

\subsection{An interesting mix}
\label{sec:an-interesting-mix}

A particular kind of asymmetric deformation is what we will call in the
following \emph{double deformation} \cite{Israel:2003cx,Kiritsis:1995iu}. At
the Lagrangian level this is obtained by adding the following marginal
perturbation to the \textsc{wzw} action:
\begin{equation}
  \delta S = \delta \kappa^2 \int \di^2 z \: J \bar J + \h \int \di^2 z \: J \bar
  I;
\end{equation}
$J$ is a holomorphic current in the group, $\bar J$ is the corresponding
anti-holomorphic current and $\bar I$ an external (to the group)
anti-holomorphic current (\emph{i.e.} in the right-moving heterotic sector
for example). A possible way to interpret this operator consists in thinking
of the double deformation as the superposition of a symmetric -- or
gravitational -- deformation (the first addend) and of an antisymmetric one
-- the electromagnetic deformation. This mix is consistent because if we
perform the $\kappa $ deformation first, the theory keeps the $U(1) × U(1) $
symmetry generated by $J$ and $\bar J$ that is needed in order to allow for
the $\h$ deformation. Following this trail, we can read off the background
fields corresponding to the double deformation by using at first one of the
methods outlined in Sec.~\ref{sec:backgr-fields-symm} and then applying the
reduction in Eq.~\eqref{eq:asym-deform} to the resulting background fields.

The final result consists in a metric, a three-form, a dilaton and a gauge
field. It is in general valid at any order in the deformation parameters $\kappa
$ and $\h$ but only at leading order in $\alpha^\prime$ due to the presence of the
symmetric part.

Double deformations of $\mathrm{AdS}_3$ where $J$ is the time-like
$J^3$ operator have been studied in~\cite{Israel:2003cx}. It was
there shown that the extra gravitational deformation allows to get
rid of the closed time-like curves, which are otherwise present in
the pure $J^3$ asymmetric deformation (Eq.~(\ref{dsnecogo})) --
the latter includes Gödel space. Here, we will  focus instead on
the case of double deformation generated by space-like operators,
$J^2$ and $\bar J^2$.

\subsection{The hyperbolic double deformation}

In order to follow the above  prescription for reading the
background fields in the double-deformed metric let us start with
the fields in Eqs.~(\ref{eq:J2J2-deform}). We can introduce those
fields in the sigma-model action. Infinitesimal variation of the
latter with respect to the parameter $\kappa^2$ enables us to
reach the following expressions for the chiral currents
$J^2_\kappa \left( z \right)$ and $\bar J^2_\kappa \left( \bar z
\right)$ at finite values of $\kappa^2$:
\begin{align}
  J^2_\kappa \left( z \right) &= \frac{1}{\cos^2 \beta + \kappa^2 \sin^2
    \beta}\left(\cos^2 \beta \: \partial \psi -\sin^2\beta
    \: \partial \varphi \right), \\
  \bar{J}^2_{\kappa} (\bar z) &= \frac{1}{\cos^2 \beta + \kappa^2 \sin^2 \beta}
  \left(\cos^2 \beta \: \partial \psi + \sin^2 \beta \: \partial \varphi
  \right).
\end{align}
Note in particular that the corresponding Killing vectors (that
clearly are $\partial_\varphi $ and $\partial_\psi $) are to be
rescaled as $L_2 = \frac {1}{\kappa^2}
\partial_\psi - \partial_\varphi $ and $R_2 = \frac {1}{\kappa^2} \partial_\psi + \partial_\varphi $. Once the
currents are known, one just has to apply the construction sketched in
Sec.~\ref{sec:backgr-fields-asymm} and write the background fields as
follows:
\begin{subequations}\label{BGDouble}
  \begin{align}
    \frac{1}{k} \di s^2 &= - \di \beta^2 + \cos^2 \beta \frac{ \left( \kappa^2 - 2
        \h^2 \right) \cos^2 \beta + \kappa^4 \sin^2 \beta }{\Delta_\kappa (\beta)^2} \di \psi^2 -
    4 \h^2 \frac{\cos^2 \beta \sin^2 \beta}{\Delta_\kappa (\beta)^2} \di \psi \di \varphi +
         \nonumber \\ & \hspace{6cm}+ \sin^2 \beta \frac{ \cos^2 \beta + \left( \kappa^2 - 2 \h^2
          \right) \sin^2 \beta}{\Delta_\kappa (\beta)^2} \di \varphi^2, \label{tmetdef} \\
    \frac{1}{k} B &= \frac{\kappa^2 - 2 \h^2 }{\kappa^2} \frac{\cos^2 \beta}{ \Delta_\kappa (\beta) } \di \varphi \land
    \di \psi,
    \\
    F &= 2 \h \sqrt{\frac{2 k}{k_g}} \frac{\sin \left( 2 \beta\right)}{\Delta_\kappa
      (\beta)^2}
    \left( \kappa^2 \di \psi \land \di \beta + \di \beta \land \di \varphi \right), \\
    \mathrm{e}^{2 \Phi } &= \frac{\sqrt{\kappa^2 - 2 \h^2}}{\Delta_\kappa
    (\beta)},
  \end{align}
\end{subequations}
where $\Delta_\kappa ( \beta ) = \cos^2 \beta + \kappa^2 \sin^2
\beta$ as in Sec.~\ref{sec:backgr-fields-symm}. In particular the
dilaton, that can be obtained by imposing the one-loop beta
equation is proportional to the ratio of the double deformed
volume form and the $\mathrm{AdS}_3 $ one.

A first observation about the above background is in order here.  The
electric field is bounded from above since $\h^2 \leq \frac{\kappa^2}{2}$. As
usual in string theory, tachyonic instabilities occur at large values of
electric or magnetic fields, and we already observed that phenomenon in Sec.
\ref{sec:backgr-fields-asymm}, for purely asymmetric ($\kappa^2 = 1$)
deformations. At the critical value of the parameter $\h$, one dimension
degenerates and the $B$-field vanishes. We are left with a two-dimensional
space (with non-constant curvature) plus electric field.

The expression~(\ref{tmetdef}) here above of the metric provides only a
local description of the space-time geometry.  To discuss the global
structure of the whole space it is useful to perform several coordinate
transformations. Firstly let us parametrize by
$\kappa^2=\lambda/(1+\lambda)$ the deformation parameter (with $\kappa <1$
for $\lambda>0$ and $\kappa >1$ for $\lambda<-1$) and introduce a radial
coordinate {\it à la } Horne and Horowitz:
\begin{equation}\label{r}
  r=\lambda +\cos^2\beta,
\end{equation}
which obviously varies between $\lambda$ and $\lambda +1$. The
expression of the metric (\ref{tmetdef}) becomes in terms of this
new coordinate:
\begin{multline}
  \di s^2 = -\left[\left( 2 \h^2 \left( 1 + \lambda \right)^2 - \lambda \right) +
      \frac{\lambda \left( \lambda - 4 \h^2 \left( 1 + \lambda \right)^2
        \right) }{r} + \frac{ 2\lambda^2 \h^2 \left( 1 + \lambda
        \right)^2))}{r^2}\right]\di \psi^2 + \\
    - \left( 1 + \lambda \right) \left[ 2 \h^2 \left( 1 + \lambda \right) +
      1 - \frac{ \left( 1 + \lambda \right) \left( 1 + 4 \h^2 \left( 1 +
            \lambda \right)^2 \right) }{r} + \frac{ 2 \left( 1 + \lambda
        \right)^3 \h^2 }{r^2} \right] \di \varphi^2 + \\
    + 4 \h^2 \left( 1 + \lambda \right)^2 \left[ 1 - \frac{ 1 + 2 \lambda
      }{r} + \frac{\lambda \left( 1 + \lambda \right) }{r^2} \right] \di \psi \di\varphi + \frac
    1{4 \left( r - \lambda \right) \left( r - \lambda - 1 \right) }\di r^2 .
\end{multline}
This expression looks close to the one discussed by Horne and
Horowitz. It also represents a black string. However, it depends
on more physical parameters as the expression of the scalar
curvature shows:
\begin{equation}
  \mathcal{R} = 2\frac{2 r \left( 1 + 2 \lambda \right) - 7  \lambda \left( 1 + \lambda
    \right) - 2  \h^2 \left( 1 + \lambda \right)^2}{r^2} .
\end{equation}

This result may seem strange at first sight since, for $\kappa=1$
and $\h=0$, the metric (\ref{tmetdef}) is of constant Ricci (and
thus scalar) curvature, corresponding to a local patch of $AdS_3$
while here, in the same limit, the curvature vanishes for large
$r$. The absence of contradiction follows from the definition of
the $r$-coordinate, becoming ill-defined for $\kappa=1$, as it
corresponds to $\lambda=\infty$.

Obviously this metric can be extended behind the initial domain of
definition of the $r$ variable. But before to discuss it, it is
interesting to note that the Killing vector ${\rm \bf
k}=(1+\lambda)\,\partial_\psi + \lambda\,\partial_\phi \propto R_2
$ is of constant square length
\begin{equation}
  \mathbf{k}.\mathbf{k} = \lambda \left( 1 + \lambda \right) - 2 \h^2
  \left( 1 + \lambda \right)^2 := \omega .
\end{equation}
Note that as $\h^2$ is positive, we have the inequality $\omega <\lambda \left( 1 +
  \lambda \right)$. Moreover, in order to have a Lorentzian signature we must impose
$\omega > 0$. The fact that the Killing vector $\mathbf{k}$ is
space-like and of constant length makes it a candidate to perform
identifications. We shall discuss this point at the end of this
section.

The constancy of the length of the Killing vector $\mathbf{k}$
suggests to make a new coordinate transformation (such that
$\mathbf{k}=\partial_x$) :
\begin{subequations} \label{tx}
  \begin{align}
    \psi &= \left( 1 + \lambda \right)  x + t ,\\
    \varphi &= t + \lambda x ,
  \end{align}
\end{subequations}
which leads to the much simpler expression of the line element:
\begin{multline}
  \label{Met2fois}
  \di s^2 = -\frac{\left( r - \lambda \right) \left( r - \lambda - 1 \right)}{r^2}\di
  t^2 + \omega \left( \di x +\frac{1}{r} \di t \right)^2 +\frac{1}{4 \left( r -
      \lambda \right) \left( r - \lambda - 1 \right)} \di r^2 .
\end{multline}
This metric is singular at $r=0, \lambda, \lambda +1$; $r=0$ being a curvature
singularity. On the other hand, the volume form is
$\nicefrac{\sqrt{\omega}}{\left(2 r \right)} \di t \land \di x \land \di r$, which
indicates that the singularities at $r=\lambda$ and $r=\lambda +1 $ may be merely
coordinate singularities, corresponding to horizons. Indeed, it is the case.
If we expand the metric, around $r=\lambda +1$, for instance, at first order
(\emph{i.e.} for $r=\lambda +1 +\epsilon$) we obtain:
\begin{equation}
  \di s^2 = \frac {\omega}{\left( 1 + \lambda \right)^2}(\di t + \left( 1 + \lambda
  \right) \di x)^2 -\frac \epsilon {\left( 1 + \lambda \right)^2} \di t \left[\di t + 2
    \frac{\omega}{1+\lambda}\left(\di t+ \left( 1 + \lambda \right)\di x\right) \right]
  +\frac 1 {4\epsilon} \di r^2
\end{equation}
indicating the presence of an horizon. To eliminate the
singularity in the metric, we may introduce Eddington--Finkelstein
like coordinates:
\begin{subequations}
  \begin{align}
    t &= \left( 1 + \lambda \right) \left (u\ ± \ \frac{1}{2} \ln \epsilon
    \right) - \omega \xi  ,    \\
    x &= \left( 1 + \frac{\omega}{ 1 + \lambda } \right) \xi - \left( u ±
      \frac{1}{2} \ln \epsilon \right) .
  \end{align}
\end{subequations}
The same analysis can also be done near the horizon located at
$r=\lambda$. Writing $r=\lambda +\epsilon$, the corresponding
regulating coordinate transformation to use is given by:
\begin{subequations}
  \begin{align}
    t &= \lambda \left( u  ±  \frac{1}{2} \ln \epsilon \right) + \omega \xi ,\\
    x &= \left( 1 - \frac{\omega}{\lambda} \right) \xi - \left( u\ ±
      \frac{1}{2} \ln \epsilon \right) .
  \end{align}
\end{subequations}
In order to reach the null Eddington--Finkelstein coordinates, we
must use null rays. The geodesic equations read, in terms of a
function $\Sigma^2[E,P,\varepsilon;r]=\left( E r - P \right)^2 -
\left( P^2/\omega \right) - \varepsilon \left( r - \lambda \right)
\left( r - \lambda - 1 \right)$:
\begin{subequations}
  \begin{align}
    \sigma &= \int \frac 1{4\, \Sigma[E,P,\varepsilon;r]}\di r,\\
    t &=\int \frac { \left(E r - P \right) r }{2 \left( r - \lambda \right)
      \left( r - \lambda - 1 \right)\, \Sigma[E,P,\varepsilon;r]} \di r   , \\
    x &= -\int \frac{\left( E r - P \right) + P / \omega}{2 \left( r -
        \lambda \right) \left( r - \lambda - 1 \right)\,
      \Sigma[E,P,\varepsilon;r]} \di r,
  \end{align}
\end{subequations}
where $E$ and $P$ are the constant of motion associated to $\partial_t$ and
$\partial_x$, $\sigma$ is an affine parameter and $\varepsilon$, equal to $1,0,-1$,
characterizes the time-like, null or space-like nature of the geodesic.
Comparing these equations (with $\varepsilon =0$ and $P=0$) with the coordinates
introduced near the horizons, we see that regular coordinates in their
neighbourhoods are given by
\begin{subequations}
  \begin{align}
    t &= T ± \frac{1}{2} \left( \left( 1 + \lambda \right) \ln\abs{ r -
        \lambda - 1 } - \lambda  \ln \abs{ r - \lambda } \right) ,\\
    x &= X \mp \frac{1}{2} \left( \ln\abs{ r - \lambda - 1 } - \ln \abs{ r - \lambda}
    \right) ,
  \end{align}
\end{subequations}
which leads to the metric
\begin{equation}
  \label{EFmet}
  \di s^2 = \left(-1+\frac{ 1 + 2 \lambda }{r} -\frac{\lambda \left( 1 + \lambda \right) -
      \omega} {r^2}\right) \di T^2 + 2 \frac \omega r \di   X \di T + \omega
  \di X^2 \mp \frac 1 r \di T  \di r.
\end{equation}
According to the sign, we obtain incoming or outgoing null
coordinates; to build a Kruskal coordinate system we have still to
exponentiate them.

Obviously, we may choose the $X$ coordinate in the metric~(\ref{EFmet}) to
be periodic without introducing closed causal curves.  The question of
performing more general identifications in these spaces will be discussed
addressed now.

We end this section by computing the conserved charges associated
to the asymptotic symmetries of our field configurations
(\ref{BGDouble}). As is well known, their expressions provide
solutions of the equations of motion derived from the low-energy
effective action
\begin{equation}
S = \int d^d x \sqrt{-g} \, \mathrm{e}^{- 2\Phi } \left[R +
4(\nabla\Phi)^2 -\frac{1}{12} H^2 -\frac{k_g}{8} F^2 +
\frac{\delta c}{3} \right]  ,
\end{equation}
in which we have choosen the units such that ${\delta c}=12$.

Expression (\ref{Met2fois}) for the metric is particularly
appropriate to describe the asymptotic properties of the solution.
In these coordinates, the various non-gravitational fields read as
\begin{eqnarray}
F&=&\pm \frac {\sqrt{2} \h (1+\lambda )}{r^2\sqrt{k_g}}\di t
\wedge \di
r ,\\
H&=& \mp \frac{\omega}{r^2}\di t\wedge \di x \wedge \di r  ,\\
\Phi&=&\Phi_\star -\frac 12 \ln r .
\end{eqnarray}
By setting $\sqrt{\omega}x=\bar x$ and $r=\rm{e}^{2\bar \rho}$,
near infinity ($\bar \rho\rightarrow \infty$), the metric
asymptotes the standard flat metric: $ds^2 = -dt^2 + d\bar{x}^2 +
d\bar \rho^2 $, while the fields $F$ and $H$ vanish and the
dilaton reads $\Phi=\Phi_\star - \bar \rho$. This allows to
interpret the asymptotic behavior of our solution (\ref{BGDouble})
as a perturbation around the solution given by $F=0$, $H=0$, the
flat metric and a linear dilaton: $\bar\Phi=\Phi_\star + f_\alpha
X^\alpha$ (here $f_\alpha=(0,0,-1)$). Accordingly, we may define
asymptotic charges associated to each asymptotic reductibility
parameter (see \cite{Glenn:2001}).

For the gauge symmetries we obtain as charges, associated to the
$H$ field
\begin{equation}
Q_H=\pm 2\rm{e}^{-2\Phi_\star}\sqrt{\omega}
\end{equation}
and to the $F$ field
\begin{equation}
Q_F=\pm \frac{2\sqrt{2}{\rm{e}}^{-2\Phi_\star}\h
(1+\lambda)}{\sqrt{k_g}} .
\end{equation}
The first one reduces (up to normalization) for $\h =0$ to the
result given in \cite{Horne:1991gn}, while the second one provides
an interpretation of the deformation parameter $\h$.

Moreover, all the Killing vectors of the flat metric defining
isometries that preserve the dilaton field allow to define
asymptotic charges. These charges are obtained by integrating the
antisymmetric tensor on the surface at infinity:
\begin{equation}
k^{[\mu \nu]}_{\xi}= {\rm{e}}^{-2\bar \Phi}\left(\xi_\sigma\
\partial_\lambda
{\cal H}^{\sigma \lambda\mu \nu} +\frac 12
\partial_\lambda\xi_\sigma\ {\cal H}^{\sigma \lambda\mu
\nu}+2\left(\xi^\mu h^\nu_\lambda f^ \lambda - \xi^\nu
h^\mu_\lambda f^\lambda\right)\right),\label{kasymp}
\end{equation}
where
\begin{equation}
{\cal H}^{\sigma \lambda\mu\nu}=\bar h^{\sigma\nu}\eta^{\lambda
\mu}+\bar h^{\lambda \mu}\eta^{\sigma\nu}-\bar
h^{\sigma\mu}\eta^{\lambda \nu}-\bar
h^{\lambda \nu}\eta^{\sigma\mu}\\
\end{equation}
is the well known tensor sharing the symmetries of the Riemann
tensor and $ \bar h^{\mu\nu}=h^{\mu\nu}-\frac 12 \eta
^{\mu\nu}\eta^{\alpha\beta}h_{\alpha\beta} $, while the Killing
vector $\xi$ has to fulfill the invariance condition $\xi_\alpha
f^\alpha = 0$. The expression of the tensor $k^{[\mu \nu]}_{\xi}$
depends only on the perturbation $h_{\mu\nu}$ of the metric tensor
because, on the one hand, the $F$ and $H$ fields appear
quadratically in the lagrangian, and their background values are
zero, while, on the other hand, the perturbation field for the
dilaton vanishes: $\Phi=\bar \Phi$ .

Restricting ourselves to constant Killing vectors, we obtain the
momenta (defined for the indice $\sigma=t$ and $\bar{x}$)
\begin{equation}\label{momenta}
P^\sigma=\int \di \bar x\ { \rm {e}}^{-2\bar
\Phi}\left(\partial_\lambda {\cal H}^{\sigma \lambda t \bar
\rho}-2 \eta ^{\sigma t} h^\nu_{\bar\rho}\right)
\end{equation}
\emph{i.e.} the density of mass ($\mu$) and momentum ($\varpi $)
per unit length:
\begin{equation}
\mu= 2{\rm{e}}^{-2\Phi_\star}(1+2\lambda)\qquad {\rm and}\qquad
\varpi=-2{\rm{e}}^{-2\Phi_\star}\sqrt{\omega}.
\end{equation}
Of course, if we perform identifications such that the string
acquires a finite length, the momenta (\ref{momenta}) become also
finite.

To make an end let us notice that the expressions of $\mu$ and
$\varpi$ that we obtain differ from those given in
\cite{Horne:1991gn} by a normalization factor but also in their
dependance with respect to $\lambda$, even in the limit $\h =0$;
indeed, the asymptotic Minkowskian frames used differ from each
other by a boost.

\section{Discrete identifications}
\label{sec:btz}

In the same spirit as the original \textsc{btz} construction reminded in
App.~\ref{btzcoo}, we would like to investigate to what extent discrete
identifications could be performed in the deformed background.
Necessary conditions for a solution~\eqref{EFmet} to remain ``viable" black
hole can be stated as follows:
\begin{itemize}
\item the identifications are to be performed along `the orbits of some
  Killing vector $\xi$ of the deformed metric
\item there must be  causally safe asymptotic regions (at spatial infinity)
\item the norm of $\xi$ has to be positive in some region of space-time, and
  chronological pathologies have to be hidden with respect to an asymptotic
  safe region by a horizon

\end{itemize}

The resulting quotient space will exhibit a black hole structure
if, once the regions where $\norm{\xi}<0$ have been removed, we
are left with an almost geodesically complete space, the only incomplete
geodesics being those ending on the locus $\norm{\xi}=0$. It is
nevertheless worth emphasizing an important difference with the
BTZ construction. In our situation, unlike the undeformed
$\mathrm{AdS}_3$ space, the initial space-time where we are to
perform identifications do exhibit curvature singularities.

\subsection{Discrete identifications in asymmetric deformations}
\label{sec:btzas}

Our analysis of the residual isometries in purely asymmetric deformations
(Sec.~\ref{sec:backgr-fields-asymm}) shows that the vector $\xi$
(Eq.~(\ref{next})) survives only in the hyperbolic deformation, whereas $\xi$
in Eq.~(\ref{ext}) is present in the parabolic one. Put differently,
non-extremal \textsc{btz} black holes allow for electric deformation, while
in the extremal ones, the deformation can only be induced by an
electro-magnetic wave.  Elliptic deformation is not compatible with
\textsc{btz} identifications.

The question that we would like to address is the following: how much of the
original black hole structure survives the deformation? The answer is
simple: a new chronological singularity appears in the asymptotic region of
the black hole. Evaluating the norm of the Killing vector shows that a naked
singularity appears. Thus the deformed black hole is no longer a viable
gravitational background. Actually, whatever the Killing vector we consider
to perform the identifications, we are always confronted to such
pathologies.

The fate of the \emph{asymmetric parabolic} deformation of $\mathrm{AdS}_3$
is similar: there is no region at infinity free of closed time-like curves
after performing the identifications.

\subsection{Discrete identifications in symmetric deformations}

Let us consider the \emph{symmetric hyperbolic} deformation, whose metric is
given by~\eqref{Met2fois} with $\h = 0$, i.e. $\omega = \lambda \left( 1 + \lambda
\right)$. This metric has two residual Killing vectors, manifestly given by
$\partial_t$ and $\partial_x$. We may thus, in general, consider identifications along
integral lines of
\begin{equation}
  \label{KillId}
  \xi=a\,\partial_t + \partial_x .
\end{equation}
This vector has squared norm:
\begin{equation}
  \norm{\xi}^2=\left( \lambda \left( 1 + \lambda \right) - a^2 \right) + \frac {2 a \lambda
    \left( 1 + \lambda \right) + a^2 \left( 1 + 2 \lambda \right) }{r} .
\end{equation}
To be space-like at infinity the vector $\xi$ must verify the
inequality $a^2 < \lambda \left( 1 + \lambda \right)$. For
definiteness, we will hereafter consider $\lambda > 0$ and $r>0$
(the case $\lambda<-1$, $r<0$ leads to similar conclusions, while
the two other situations have to be excluded in order to avoid
naked singularities, see eq.~(\ref{Met2fois})). If $a > 0 $, or
$-\sqrt{ \lambda \left( 1 + \lambda
  \right) } < a < -2\lambda \left( 1 + \lambda \right) / \left( 1 + 2 \lambda \right)$,
$\xi$ is everywhere space-like. Otherwise, it becomes time-like
behind the inner horizon ($r = \lambda$), or on this horizon if
$a=-\lambda$. In this situation, the quotient space will exhibit a
structure similar to that of the black string, with a time-like
chronological singularity (becoming light-like for $a=-\lambda$)
hidden behind two horizons (or a single one for $a=-\lambda$).

\subsection{Discrete identifications in double deformations}
\label{sec:btz-ident-comb}

The norm squared of the identification vector (\ref{KillId}) in the
metric~\eqref{Met2fois} is
\begin{equation}
  \norm{\xi}^2= \left( \omega - a^2 \right) + 2 \frac{a \omega +
    a^2 \left( 1 + 2 \lambda \right) }{r} - \frac{a^2 \left( \lambda \left( 1+\lambda
      \right) - \omega \right)}{r^2} .
\end{equation}
Between $r=0$ and $r=\infty$, this scalar product vanishes once
and only once (if $a\neq 0$).  To be space-like at infinity we
have to restrict the time component of $\xi$ to $\abs{a} < \omega
$. Near $r=0$ it is negative, while near the inner horizon
($r=\lambda$) it takes the non-negative value $\omega \left(
\lambda + a \right)^2/\lambda^2$. Accordingly, by performing
identifications using this Killing vector, we will encounter a
chronological singularity, located at $r=r^*$, with
$0<r^*\leq\lambda$,  the singularity being of the same type as the
one in the symmetric case (see  Fig.~\ref{Penrose}).

\FIGURE{\includegraphics[width=9cm]{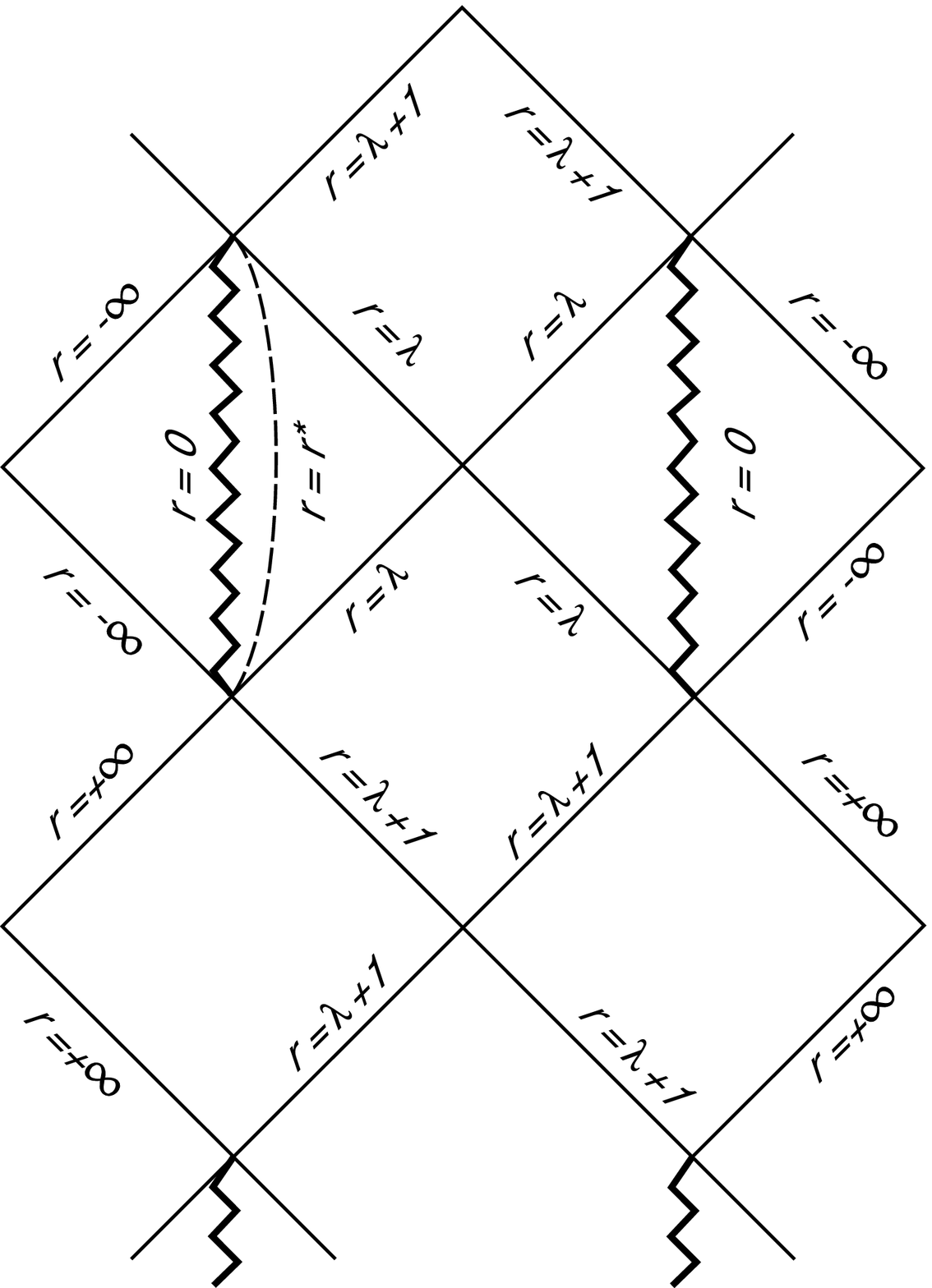}

  \caption{Penrose diagram exhibiting the global structure of the
    double hyperbolic deformation. The time-like curvature singularities
    $r=0$ are represented, as well as the horizons, located at $r=\lambda$ and
    $r=\lambda +1$. When performing identifications along orbits of Killing
    vectors that allow for a causally safe region at infinity,
    a time-like chronological singularity may appear at $r=r^*$, with $0 < r^*
\leq \lambda$.}
 \label{Penrose}}










\section{Towards the exact spectra}
\label{sec:conf-field-theory}

The main guideline for exploring the black hole geometries that we have so
far considered has been the presence of an underlying \textsc{cft}
description. This allows us to identify the background fields as the
Lagrangian counterparts of exact conformal field theories. In this section
we will give a look to the other -- algebraic -- aspect of these models,
showing how it is possible to write an explicit expression for the spectrum
of primary operators. 

Since this kind of contruction has already been carried on
in~\cite{Israel:2003cx} for the $J_3 $ double deformation of $\SL$, here we
will focus of the $J_2$ deformations, giving the spectrum for the deformed
theory (Sec.~\eqref{sec:deformed-spectrum}) and for a deformed theory with
discrete identifications (Sec.~\eqref{sec:twisting}). We will limit
ourselves to giving the spectrum for the theory: the evaluation of the
partition function, although straightforward in principle, would require the
decomposition of the $\SL$ partition function in a hyperbolic basis of
characters, a still unresolved problem.

\subsection{Deformed Spectrum}
\label{sec:deformed-spectrum}

Consider the double deformation described in Sec.~\ref{sec:two-parameter}
for a $SL (2, \setR )_k$ super-\textsc{wzw} model where $J$ is the
hyperbolic (space-like) $J_2$ current.

The evaluation of the spectrum for our deformed model is pretty
straightforward once one realizes that the deformations act as $O \left( 2,2
\right)$ pseudo-orthogonal transformations on the charge lattice
corresponding to the abelian subgroup of the $\mathfrak{sl}(2,\setR)$
heterotic model (as described in Sec.~\ref{sec:wzw-deformations}). Left and
right weights for the relevant lattices are (see
Eqs.~\eqref{eq:left-weights} and~\eqref{eq:right-weights}):
\begin{subequations}
  \begin{align}
    L_0 &= \frac{1}{k} \left( \mu +n + \frac{a}{2} \right)^2, \\
    \bar L_0 &= \frac{\bar \mu^2}{k+2} + \frac{1}{k_g} \left( \bar n +
\frac{\bar a}{2} \right)^2,
  \end{align}
\end{subequations}
where the anti-holomorphic part contains the contribution coming from a
$\mathfrak{u}(1)$ subgroup of the heterotic gauge group.

At the Lagrangian level, the infinitesimal deformation we want to describe is
given by the following marginal operator:
\begin{equation}
  \mathcal{O} = \kappa^2  \frac{\left(J^2 + \imath \psi_1 \psi_3
\right)}{\sqrt{k}} \frac{\bar J^2}{\sqrt{k+2}} + \h
\frac{\left(J^2 + \imath \psi_1 \psi_3 \right)}{\sqrt{k}}
\frac{\bar I}{\sqrt{k_g}}.
\end{equation}
This suggests that the actual $O (2,2)$ transformation should be
obtained as a boost between the holomorphic part and the result of
a rotation between the two anti-holomorphic components. The
deformed lattices then read:
\begin{subequations}
  \begin{align}
    L_0^{\text{dd}} &= \Set{\frac{1}{\sqrt{k}} \left( \mu + n + \frac{a}{2} \right) \cosh x
+ \left( \frac{\bar \mu }{\sqrt{k+2}} \cos \alpha +
\frac{1}{\sqrt{k_g}} \left( \bar n + \frac{\bar a}{2} \right) \sin
\alpha \right) \sinh x }^2,
\label{eq:rotation-left}\\
    \bar L_0^{\text{dd}} &= \Set{  \left( \frac{\bar \mu }{\sqrt{k+2}} \cos \alpha +
\frac{1}{\sqrt{k_g}} \left( \bar n + \frac{\bar a}{2} \right) \sin
\alpha \right) \cosh x + \frac{1}{\sqrt{k}} \left( \mu + n +
\frac{a}{2} \right) \sinh x }^2, \label{eq:rotation-right}
  \end{align}
\end{subequations}
where the parameters $x $ and $\alpha$ can be expressed as
functions of $\zeta$ and $\xi $ as follows:
\begin{equation}
  \begin{cases}
    \kappa^2 = \sinh (2x) \cos \alpha, \\
    \h = \sinh (2x) \sin \alpha.
  \end{cases}
\end{equation}

\subsection{Twisting}
\label{sec:twisting}

The identification operation we performed in the symmetrically and
double-deformed metric (as in Sec.~\ref{sec:btz}) is implemented in the
string theory framework by the orbifold construction. This was already
obtained in~\cite{Natsuume:1998ij,Hemming:2001we} for the ``standard''
\textsc{btz} black hole that was described as a $SL (2, \setR )/ \setZ$
orbifold.

In order to write the spectrum that will contain the twisted sectors, the first
step consists in writing explicitly the primary fields in our theory,
distinguishing between the holomorphic and anti-holomorphic parts (as it is
natural to do since the construction is intrinsically heterotic).
\begin{itemize}
\item The holomorphic part is simply written by introducing the charge boost of
Eq.~\eqref{eq:rotation-left} in Eq.~\eqref{eq:superparafermion}:
  \begin{equation}
    \Phi^{\text{dd}}_{j \mu \nu \bar \mu \bar \nu} (z) = U_{j \mu } (z) \exp \left[  \imath \left( \sqrt{\frac{2}{k}} \left( \mu + n + \frac{a}{2} \right) \cosh x +
    \sqrt{2} \bar Q_\alpha \sinh x \right) \vartheta_2 \right],
  \end{equation}
  where $Q_\alpha = \bar \mu \sqrt{\frac{2}{k+2}} \cos \alpha + \bar \nu
  \sqrt{\frac{2}{k_g}} \sin \alpha$ and the dd superscript stands for double
  deformed
\item To write the anti-holomorphic part we need at first to implement the
rotation between the $\bar J^3$ and gauge current components:
  \begin{multline}
    \bar \Phi_{j \bar \mu \bar \nu} ( \bar z ) = V_{j \mu } (\bar z) {\rm
      e}^{\imath \bar \mu \sqrt{\nicefrac{2}{k+2}} \bar \theta_2} {\rm
      e}^{\imath \bar \nu \sqrt{2/k_g}
      \bar X}  = \\
    = V_{j \mu } (\bar z) {\rm e}^{\imath \sqrt{2} \bar Q_{\alpha} \left(
        \bar \theta_2 \cos \alpha + \bar X \sin \alpha \right) } {\rm
      e}^{\imath \sqrt{2} \bar Q_{\alpha - \nicefrac{\pi}{2}} \left( - \bar
        \theta_2 \sin \alpha + \bar X \cos \alpha \right) },
  \end{multline}
  and then realize  the boost in Eq.~\eqref{eq:rotation-right} on the involved part:
  \begin{multline}
    \bar \Phi^{\text{dd}}_{j \mu \bar \mu \nu \bar \nu} ( \bar z ) = V_{j
      \mu } {\rm e}^{\imath \sqrt{2} \bar Q_{\alpha - \nicefrac{\pi}{2}}
      \left( - \bar \theta_2
        \sin \alpha + \bar X \cos \alpha \right)}  \times \\
    \times \exp \left[ \imath \left( \sqrt{\frac{2}{k}} \left( \mu + n +
          \frac{a}{2} \right) \sinh x + \sqrt{2} \bar Q_\alpha \cosh x
      \right) \left( \bar \theta_2 \cos \alpha + \bar X \sin \alpha \right)
    \right].
  \end{multline}
\end{itemize}

Now that we have the primaries, consider the operator $W_w \left( z, \bar z
\right)$ defined as follows:
\begin{equation}
  W_w \left( z, \bar z \right) = {\rm e}^{-\imath \frac{k}{2} w \Delta_-
    \vartheta_2 +\imath \frac{k+2}{2} w\Delta_+ \bar \theta_2 },
\end{equation}
where $w \in \setZ $ and $\bar \theta_2 $ the boson corresponding to the
$\bar J_2 $ current. It is easy to show that the following \textsc{ope}'s
hold:
\begin{align}
  \vartheta_2 \left( z \right) W_n \left( 0, \bar z\right) &\sim -\imath w
\Delta_- \log z
  W_w
  \left( 0, \bar z\right), \\
  \bar \theta_2 \left( \bar z \right) W_n \left( z, 0 \right) &\sim \imath w
  \Delta_+ \log \bar z W_w \left( z, 0\right),
\end{align}
showing that $W_w \left( z, \bar z\right)$ acts as twisting operator with
winding number $w$ ($\vartheta_2 $ and $\bar \theta_2 $ shift by $2 \pi
\Delta_- w $ and $2 \pi \Delta_+ w $ under $z \to {\rm e}^{2 \pi \imath } z
$). This means that the general primary field in the $SL \left( 2, \setR
\right)_k / \setZ $ theory can be written as:
\begin{equation}
  \Phi^{\text{tw}}_{j \mu \bar \mu \nu \bar \nu  w} \left( z, \bar z \right) =  \Phi^{\text{dd}}_{j \mu \bar  \mu \nu \bar \nu  } \left( z, \bar z \right) W_w \left( z, \bar z 
    \right).
\end{equation}
where the tw superscript stands for twisted.

Having the explicit expression for the primary field, it is simple to derive
the scaling dimensions which are obtained, as before, via the \textsc{gko}
decomposition of the Virasoro algebra $T \left[ \mathfrak{sl}\left( 2, \setR
  \right) \right] = T \left[ \mathfrak{sl}\left( 2, \setR
  \right)/\mathfrak{o} \left( 1,1\right) \right] + T \left[ \mathfrak{o}
  \left(1,1 \right) \right] $. Given that the $ T \left[ \mathfrak{sl}\left(
    2, \setR \right)/\mathfrak{o} \left( 1,1\right) \right] $ part remains
invariant (and equal to $L_0 = - j \left( j+1\right)/ k - \mu^2/\left( k +
  2\right)$ as in Eq.~\eqref{eq:boson-coset}), the deformed weights read:
\begin{subequations}
  \begin{align}
    L^{\text{tw}}_0 &= \Set{ \frac{k}{2\sqrt{2}} w \Delta_-
    + \frac{1}{\sqrt{k}} \left( \mu + n + \frac{a}{2} \right) \cosh x
+  \bar Q_\alpha \sinh x }^2,\\
    \bar L^{\text{tw}}_0 &= \Set{ - \frac{k+2}{2\sqrt{2}} w \Delta_+ \cos
      \alpha+ \bar Q_\alpha \cosh x + \frac{1}{\sqrt{k}}
      \left( \mu + n + \frac{a}{2} \right) \sinh x }^2 + \nonumber \\
    & \hspace{6.5cm}+ \Set{ \frac{k+2}{2 \sqrt{2}} w \Delta_+ \sin \alpha + \bar Q_{\alpha -
        \nicefrac{\pi}{2}}}^2.
  \end{align}
\end{subequations}


\section{Summary}

The main motivation for this work has been a systematic search of
black-hole structures in the moduli space of AdS$_3$, via marginal
deformations of the $\SL$ \textsc{wzw} model and discrete
identifications. This allows to reach three-dimensional geometries
with black-hole structure that generalize backgrounds such as the
\textsc{btz} black hole \cite{Banados:1992wn} or the
three-dimensional black string \cite{Horne:1991gn}.

The backgrounds under consideration include a (singular) metric, a
Kalb--Ramond field, a dilaton and an electric field. The latter is
always bounded from above, as usual in string theory, where
tachyonic instabilities are expected for large electric or
magnetic fields.

We have computed parameters such as mass or charge. For
backgrounds obtained by performing marginal deformations, those
parameters are related to the deformation parameters.
Singularities are true curvature singularities, hidden behind
horizons. This is to be opposed to the \textsc{btz} black-holes,
where masses and momenta are introduced by the Killing vector of
the discrete identification, and where the singularity is a
chronological singularity.

Discrete identifications \emph{à la} \textsc{btz} can be
superimposed to the black holes obtained by continuous
deformations of AdS$_3$. Extra chronological singularities appear
in that case, which force us to excise some part of the original
space. This part turns out to contain the locus of the curvature
singularity. It is worth stressing that for certain range of the
deformation parameters, \emph{naked} singularities appear.

Although the geometrical view point has been predominating, the
guideline for our study comes from the underlying \textsc{cft}
structure. This has enabled us to provide both a geometrical and
an algebraical description in terms of the spectrum of the string
primaries.

Since we are dealing with the extension of $\mathrm{AdS}_3$ one
may wonder about a possible holographic interpretation for the
exact string backgrounds at hand, aiming at generalizing the usual
$\mathrm{AdS}$/\textsc{cft} correspondence. A major obstruction to
this is due to the asymptotic flatness of the geometries. Hence,
it is not clear how to find a suitable boundary map.

Interesting questions that we did not address, which are in
principle within reach, are those dealing with the thermodynamical
properties of the above black holes, for which a microscopic
interpretation in terms of string states should be tractable.


\acknowledgments

D.O. and M.P. thank D. Israël, E. Kiritsis, C. Kounnas and K.
Sfetsos for useful discussions. S.D. and Ph.S. are grateful to G.
Compère for its patient explanation of the general method for
computing charges; S.D. thanks P. Aliani and D. Haumont for their
advice. S.D. and Ph.S. acknowledge support from the Fonds National
de la Recherche scientifique through an F.R.F.C. grant. D.O. and
M.P. thank UMH and ULB for kind hospitality at various stages of
the project, and acknowledge support from the E.U. under the
contracts \textsc{mext-ct}-2003-509661,
\textsc{mrtn-ct}-2004-005104 and \textsc{mrtn-ct}-2004-503369.

\appendix

\section{\boldmath $\mathrm{AdS}_3$ coordinate patches\unboldmath}
\label{antids}

\subsection{$\mathrm{AdS}_3$ $SL(2, \mathbb{R})$}
\label{sl2}

The commutation relations for the generators of the $SL(2,\mathbb{R})$
algebra are
\begin{align}
  \comm{ J^1 , J^2} = - \imath J^3 && \comm{ J^2 , J^3} = \imath J^1 &&
  \comm{ J^3 , J^1 } = \imath J^2.
  \label{comm}
\end{align}
The three-dimensional anti-de Sitter space is the universal
covering of the $SL(2,\mathbb{R})$ group manifold. The latter can
be embedded in a Lorentzian flat space with signature $(-,+,+,-)$
and coordinates $(x^0,x^1,x^2,x^3)$:
\begin{equation}
  g  =  L^{-1}\
  \begin{pmatrix}
    x^0 + x^2 & x^1 + x^3 \\ x^1 -
      x^3 & x^0 - x^2
  \end{pmatrix}, \label{4emb}
\end{equation}
where $L$ is the radius of $\mathrm{AdS}_3$.

The isometry group of the $SL(2,\mathbb{R})$ group manifold is
generated by left or right actions on $g$: $g\to hg$ or $g\to gh$
$\forall h \in SL(2,\mathbb{R})$. From the four-dimensional point
of view, it is generated by the Lorentz boosts or rotations
$\zeta_{ab}= i\left( x_a\partial_b - x_b
  \partial_a\right)$ with $x_a=\eta_{ab}x^b$. We list here explicitly the
six Killing vectors, as well as the group action they correspond
to:
\begin{subequations}
  \label{eq:Killing-SL2R}
  \begin{align}
    L_1 &= \frac{\imath }{2}\left(\zeta_{32} - \zeta_{01}\right), & g &\to
    \mathrm{e}^{-\frac{\lambda}{2}\sigma^1}g,
    \label{L1} \\
    L_2 &= \frac{\imath }{2}\left(-\zeta_{31}-\zeta_{02} \right), &  g &\to
    \mathrm{e}^{-\frac{\lambda}{2}\sigma^3}g,
    \label{L2} \\
    L_3 &= \frac{\imath }{2}\left(\zeta_{03} - \zeta_{12}\right), & g&\to
    \mathrm{e}^{\imath\frac{\lambda}{2}\sigma^2}g,
    \label{L3} \\
    R_1 &= \frac{\imath }{2}\left( \zeta_{01} + \zeta_{32}\right), & g&\to
    g\mathrm{e}^{\frac{\lambda}{2}\sigma^1},
    \label{R1} \\
    R_2 &= \frac{\imath }{2}\left(\zeta_{31} - \zeta_{02}\right), &  g&\to
    g\mathrm{e}^{-\frac{\lambda}{2}\sigma^3},
    \label{R2} \\
    R_3 &= \frac{\imath }{2}\left(\zeta_{03} + \zeta_{12}\right), & g&\to
    g\mathrm{e}^{\imath\frac{\lambda}{2}\sigma^2}.
    \label{R3}
  \end{align}
\end{subequations}

Both sets satisfy the algebra (\ref{comm}) (once multiplied by
$-\imath $). The norms of the Killing vectors are the following:
\begin{equation}
  \norm{L_1}^2 = \norm{R_1}^2 = \norm{L_2}^2 =
  \norm{R_2}^2 =- \norm{L_3}^2=-\norm{R_3}^2 = \frac{L^2}{4}.
\end{equation}
Moreover $L_i \cdot L_j = 0$ for $i\neq j$ and similarly for the
right set. Left vectors are not orthogonal to right ones.

The isometries of the $SL(2,\mathbb{R})$ group manifold turn into
symmetries of the $SL(2,\mathbb{R})_k$ \textsc{wzw} model, where
they are realized in terms of conserved currents\footnote{When
writing actions a choice of gauge
  for the \textsc{ns} potential is implicitly made, which breaks part of the
  symmetry: boundary terms appear in the transformations.  These must be
  properly taken into account in order to reach the conserved currents.
  Although the expressions for the latter are not unique, they can be put in
  an improved-Noether form, in which they have only holomorphic (for
  $L_i$'s) or anti-holomorphic (for $R_j$'s) components.}.
The reader will find details on those issues in the appendices of
\cite{Israel:2004vv}.

\subsection{``Symmetric" coordinates} \label{antidss}

One introduces Euler-like angles by
\begin{equation}
g = \mathrm{e}^{\imath {\tau+\phi \over 2} \sigma^2}
\mathrm{e}^{\rho \sigma^1} \mathrm{e}^{\imath{\tau-\phi \over 2}
\sigma^2} , \label{euler}
\end{equation}
which provide good global coordinates for $\mathrm{AdS}_3$ when
$\tau\in ]-\infty,+\infty[$, $\rho\in [0,\infty[$, and $\phi\in
[0,2\pi]$. In Euler angles, the invariant metric reads:
\begin{equation}
\di s^2= L^2\left[- \cosh ^2 \rho  \, \di \tau ^2 +\di \rho^2 +
\sinh^2 \rho \, \di \phi^2\right]. \label{dseul}
\end{equation}
The Ricci scalar of the corresponding Levi--Civita connection is
$\mathcal{R}=-6/L^2$. The volume form reads:
\begin{equation}
  \label{vfeul}
  \omega_{[3]} = \frac{L^3}{2}\sinh 2\rho
     \di \rho \land \di \phi  \land \di \tau,
\end{equation}
whereas $L_3 = \frac{1}{2}\left( \d_\tau + \d_\phi\right)$ and
$R_3 = \frac{1}{2}\left( \d_\tau - \d_\phi\right)$.

Another useful, although not global, set of coordinates is defined
by
\begin{equation}
  g = {\rm e}^{{\frac{\psi - \varphi}{2}} \sigma^3} {\rm e}^{\imath \beta \sigma^1}
  {\rm e}^{\frac{\psi + \varphi }{2} \sigma^3},\label{sphan}
\end{equation}
($\psi $ and $\varphi$ \emph{are not} compact coordinates).  The metric reads:
\begin{equation}
  \di s^2 = L^2\left[ \cos ^2 \beta  \di \psi^2 -\di \beta^2 + \sin^2 \beta\,
    \di \varphi^2\right], \label{dssphan}
\end{equation}
with volume form
\begin{equation}
  \label{vfsphan}
  \omega_{[3]} = \frac{L^3}{2}\sin 2\beta \di \beta \land \di \psi  \land \di \varphi.
\end{equation}
Now $L_2 = \frac{1}{2}\left( \d_\psi - \d_\varphi\right)$ and $R_2 =
\frac{1}{2}\left( \d_\psi + \d_\varphi \right)$.

Finally, the Poincaré coordinate system is defined by
  \begin{equation} \label{eq:ads-poinc-tr}
    \begin{cases}
    x^0 + x^2 &=\frac{L}{u},\\
    x^0 - x^2 &= Lu + \frac{L x^+ x^-}{u},\\
    x^1 ± x^3 &= \frac{L x^±}{u} .
    \end{cases}
  \end{equation}
  For $\set{u,x^+,x^-} \in \mathbb{R}^3$, the Poincaré
  coordinates cover once the $SL(2\mathbb{R})$ group manifold. Its
  universal covering, $\mathrm{AdS}_3$, requires an infinite
  number of such patches. Moreover, these coordinates exhibit
  a Rindler horizon at $\vert u \vert \to \infty$; the conformal
  boundary is at $\vert u \vert \to 0$.
  Now the metric reads:
  \begin{equation}
    \di s^2 = \frac{L^2}{u^2} \left( \di u^2 + \di x^+ \di x^-
    \right),
  \end{equation}
  and the volume form:
  \begin{equation}
    \label{eq:ads-poinc-volume}
    \omega_{[3]} = \frac{L^3}{2u^3} \di u \land \di x^+ \land \di
    x^-.
  \end{equation}
We also have $L_1+L_3 = -\d_-$ and $R_1+R_3 = \d_+$.

\subsection{``Asymmetric" coordinates} \label{antidsas}

The above three sets of AdS$_3$ coordinates are suitable for
implementing symmetric parabolic, elliptic or hyperbolic
deformations, respectively driven by $\left( J^1 +J^3
\right)\left(\bar J^1 + \bar J^3 \right)$, $J^3\bar J^3$ or  $J^2
\bar J^2$. For asymmetric elliptic or hyperbolic deformations, we
must use different coordinate systems, where the structure of
$\mathrm{AdS}_3 $ as a Hopf fibration is more transparent. They
are explicitly described in the following.
\begin{itemize}
\item The coordinate system used to describe
  the elliptic asymmetric deformation is defined as follows:
  \newcommand{\CR}[0]{\cosh \frac{\rho}{2}}
  \newcommand{\SR}[0]{\sinh \frac{\rho}{2}}
  \newcommand{\CPHI}[0]{\cosh \frac{\phi}{2}}
  \newcommand{\SPHI}[0]{\sinh \frac{\phi}{2}}
  \newcommand{\CT}[0]{\cos \frac{t}{2}}
  \newcommand{\ST}[0]{\sin \frac{t}{2}}
  \begin{equation}
    \begin{cases}
      \frac{x_0}{L} &= \CR \CPHI \CT - \SR \SPHI \ST, \\
      \frac{x_1}{L} &= -\SR \SPHI \CT - \CR \SPHI \ST,\\
      \frac{x_2}{L} &= -\CR \SPHI \CT + \SR \CPHI \ST, \\
      \frac{x_3}{L} &= -\SR \SPHI \CT - \CR \CPHI \ST.
    \end{cases}
  \end{equation}
  The metric now reads:
   \begin{equation}
     \label{eq:ads-rhotphi-metric}
     \di s^2 = \frac{L^2}{4} \left( \di \rho^2 + \di \phi^2 - \di t^2 -
       2 \sinh \rho \di t \di \phi\right),
   \end{equation}
   and the corresponding volume form is
   \begin{equation}
     \label{eq:ads-rhotphi-vf}
     \omega_{[3]} = \frac{L^3}{8}\cosh \rho
     \di \rho \land \di \phi  \land \di t.
  \end{equation}
 This coordinate
  system is such that the $t$-coordinate lines coincide with the integral
  curves of the Killing vector $L_3 = - \d_t$, whereas the $\phi$-lines are
  the curves of $R_2 = \d_\phi$.
\item The coordinate system used to describe the asymmetric
  hyperbolic deformation is defined as follows:
  \renewcommand{\CR}[0]{\cosh \frac{r}{2}}
  \renewcommand{\SR}[0]{\sinh \frac{r}{2}}
  \newcommand{\CX}[0]{\cosh \frac{x}{2}}
  \newcommand{\SX}[0]{\sinh \frac{x}{2}}
  \renewcommand{\CT}[0]{\cos \frac{\tau}{2}}
  \renewcommand{\ST}[0]{\sin \frac{\tau}{2}}
  \begin{equation}
    \label{eq:ads-rxt-coo}
    \begin{cases}
      \frac{x_0}{L} &= \CR \CX \CT + \SR \SX \ST, \\
      \frac{x_1}{L} &= - \SR \CX \CT + \CR \SX \ST, \\
      \frac{x_2}{L} &= - \CR \SX \CT - \SR \CX \ST, \\
      \frac{x_3}{L} &= \SR \SX \CT - \CR \CX \ST.
    \end{cases}
  \end{equation}
  For $\set{r,x,\tau} \in \mathbb{R}^3$, this patch covers exactly once the
  whole $\mathrm{AdS}_3$, and is regular
  everywhere~\cite{Coussaert:1994tu}.  The metric is then given by
  \begin{equation}
     \label{eq:ads-rxt-met}
    \di s^2 = \frac{L^2}{4} \left( \di r^2 + \di x^2 - \di \tau^2 +
      2 \sinh r \di x \di \tau \right),
  \end{equation}
  and correspondingly the volume form is
  \begin{equation}
    \label{eq:ads-rxt-vf}
    \omega_{[3]} = \frac{L^3}{8} \cosh r \di r \land \di x \land \di \tau .
  \end{equation}
  In this case the $x$-coordinate lines
  coincide with the integral curves of the Killing vector $L_2 = \d_x$,
  whereas the $\tau$-lines are the curves of $R_3=-\d_\tau$.
\end{itemize}
\section{\boldmath The  BTZ black hole\unboldmath}
\label{btzcoo}

In the presence of isometries, discrete identifications provide
alternatives for creating new backgrounds. Those have the
\emph{same} local geometry, but differ with respect to their
global properties. Whether these identifications can be
implemented as orbifolds at the level of the underlying
two-dimensional string model is very much dependent on each
specific case.

For $\mathrm{AdS}_3$, the most celebrated geometry obtained by
discrete identification is certainly the \textsc{btz} black
hole~\cite{Banados:1992wn}. The discrete identifications are made
along the integral lines of the following Killing vectors (see
Eqs. (\ref{eq:Killing-SL2R})):
\begin{subequations}
  \begin{align}
     \text{non-extremal case}&: \ \ \xi =
     \left(r_+ + r_- \right)R_2 -  \left(r_+ - r_- \right)L_2,\label{next}\\
    \text{extremal case}&: \ \ \xi = 2 r_+  R_2 -
     \left( R_1-R_3\right) -
      \left( L_1+L_3\right) \label{ext}.
  \end{align}
\end{subequations}

In the original \textsc{btz} coordinates,
the metric reads:
\begin{equation}
\di s^2 =L^2 \left[ -f^2(r) \, \di t^2 + f^{-2}(r) \, \di r^2 +
r^2 \left( \di \varphi-\frac{r_+  r_-}{r^2} \di t
\right)^2\right] ,
\end{equation}
with
\begin{equation}
f(r) =\frac{1}{r}\sqrt{\left( r^2_{\vphantom +}-r^2_{+} \right)
\left(r^2_{\vphantom -}-r^2_{-}\right)} .
\end{equation}
In this coordinate system,
\begin{equation}
\partial_{\varphi} \equiv \xi \ , \ \
\partial_t \equiv
-\left(r_+ + r_- \right) R_2 - \left(r_+ - r_- \right) L_2\ \ {\rm
and} \ \ r^2 \equiv \norm{\xi}.
\end{equation}
In AdS$_3$ $\varphi$ is not a compact coordinate. The discrete
identification makes $\varphi$ an angular variable, $\varphi \cong
\varphi+2\pi$, which imposes to remove the region with $r^2<0$.
The \textsc{btz} geometry describes a three-dimensional black
hole, with mass $M$ and angular momentum $J$, in a space--time
that is locally (and asymptotically) anti-de Sitter. The
chronological singularity at $r=0$ is hidden behind an inner
horizon at $r=r_-$, and an outer horizon at $r=r_+$. Between these
two horizons,  $r$ is time-like. The coordinate $t$ becomes
space-like inside the ergosphere, when $r^2_{\vphantom g}<
r^2_{\rm erg} \equiv r_+^2 + r_-^2$. The relation between $M,J$
and  $r_±$ is as follows:
\begin{equation}
r_±^2  = {ML \over 2}\left[ 1± \sqrt{1-\left({J\over  ML
}\right)^2} \right] .
\end{equation}
Extremal black holes have  $\vert J \vert =ML$ ($r_+ = r_-$). In
the special case $J= ML=0$ one finds the near-horizon geometry of
the five-dimensional  \textsc{ns5/f1} stringy black hole in its
ground state. Global  AdS$_3$  is obtained for $J=0$ and $ML=-1$.

Many subtleties arise, which concern \emph{e.g.} the appearance of
closed time-like curves in the excised region of negative  $r^2$
(where $\partial_\varphi$ would have been time-like) or the
geodesic completion of the manifold; a comprehensive analysis of
these issues can be found in~\cite{Banados:1993gq}. At the
string-theory level, the \textsc{btz} identification is realized
as an \emph{orbifold} projection, which amounts to keeping
invariant states and adding twisted
sectors~\cite{Natsuume:1998ij,Hemming:2001we}.

Besides the \textsc{btz} solution, other locally $\mathrm{AdS}_3$
geometries are obtained, by imposing identification under purely
left (or right) isometries, refereed to as self-dual (or
anti-self-dual) metrics. These were studied
in~\cite{Coussaert:1994tu}. Their classification and isometries
are exactly those of the asymmetric deformations studied in the
present chapter. The Killing vector used for the identification is
(A) time-like (elliptic), (B) space-like (hyperbolic) or (C) null
(parabolic), and the isometry group is $U(1) ×
SL(2,\mathbb{R})$. It was pointed out in~\cite{Coussaert:1994tu}
that the resulting geometry was free of closed time-like curves
only in the case (B).

\section{\boldmath Spectrum of the $ \SL$ super-\textsc{wzw} model \unboldmath}
\label{sec:initial-spectrum}

In this appendix we give a reminder of the superconformal \textsc{wzw} model
on $SL \left( 2, \setR \right)_k$ (for a recent discussion see
\cite{Giveon:2003wn}). The affine extension of the $\mathfrak{sl} \left( 2,
  \setR \right)$ algebra at level $k$ is obtained by considering two sets of
holomorphic and anti-holomorphic currents of dimension one, defined as
\begin{align}
  J^{\textsc{m}} \left( z \right) = k \braket{T^{\textsc{m}}, \mathrm{Ad}_g
    g^{-1} \partial g}, && \bar J^{\textsc{m}} \left( \bar z \right) = k
  \braket{T^{\textsc{m}}, g^{-1} \bar \partial g},
\end{align}
where $\braket{\cdot, \cdot } $ is the scalar product (Killing
form) in $\mathfrak{sl} \left( 2, \setR \right)$,
$\set{T^{\textsc{m}}} $ is a set of generators of the algebra that
for concreteness we can choose as follows:
\begin{align}
  T^1 = \sigma^1, && T^2 = \sigma^3, && T^3 = \sigma^2.
\end{align}
Each set satisfies the \textsc{ope}
\begin{equation}
  J^{\textsc{m}} \left( z \right) J^{\textsc{n}} \left( w \right) \sim \frac{k
    \delta^{\textsc{mn}}}{2 \left( z - w\right)^2} +
  \frac{f^{\textsc{mn}}_{\phantom{\textsc{mn}}\textsc{p}} J^{\textsc{p}}
    \left( w \right)}{z-w},
\end{equation}
where $f^{\textsc{mn}}_{\phantom{\textsc{mn}}\textsc{p}}$ are the
structure constants of the $\mathfrak{sl} \left( 2, \setR \right)$
algebra.  The chiral algebra contains the Virasoro operator
(stress tensor) obtained by the usual Sugawara construction:
\begin{equation}
  T \left( z \right) = \sum_{\textsc{m}} \frac{: J^{\textsc{m}} J^{\textsc{m}}
    :  }{k-2}.
\end{equation}

A heterotic model is built if we consider a left-moving
$\mathcal{N} = 1 $ extension, obtained by adding 3
free fermions which transform in the adjoint representation. More
explicitly:
\begin{align}
  T \left( z \right) &= \sum_{\textsc{m}} \frac{: J^{\textsc{m}} J^{\textsc{m}}
    :  }{k-2} + : \psi_{\textsc{m}} \partial \psi_{\textsc{m}}:, \\
  G \left( z \right) &= \frac{2}{k} \left( \sum_{\textsc{m}} J^{\textsc{m}}
    \psi_{\textsc{m}} - \frac{\imath }{3k } \sum_{\textsc{mnp}} f^{\textsc{mnp}}:
    \psi_{\textsc{m}} \psi_{\textsc{n}} \psi_{\textsc{p}} :
    \right).
\end{align}
On the right side, instead of superpartners, we add a right-moving current
with total central charge $c=16$.

Let us focus on the left-moving part. The supercurrents are given by
$\psi_{\textsc{m}} + \theta \sqrt{2/k} \mathcal{J}_{\textsc{m}}$ where:
\begin{equation}
  \mathcal{J}_{\textsc{m}} = J^{\textsc{m}} - \frac{\imath }{2}
\sum_{\textsc{np}}\epsilon^{\textsc{mnp}}
  \psi_{\textsc{n}} \psi_{\textsc{p}};
\end{equation}
it should be noted that the bosonic $J^{\textsc{m}} $ currents generate an
affine $\mathfrak{sl}\left( 2, \setR \right)$ algebra at level $k+2$, while
the level for the total $\mathcal{J}_{\textsc{m}}$ currents is $k$.

Let us now single out the operator that we used for both the deformation
(Eqs.~\eqref{tmetdef}) and the identifications
(Sec.~\ref{sec:btz-ident-comb}):
\begin{equation}
  \mathcal{J}_2 = J^2 + \imath \psi_1 \psi_3.
\end{equation}
Let us now bosonize these currents as follows:
\begin{align}
  \mathcal{J}_2 &= - \sqrt{ \frac{k}{2}} \partial \vartheta_2,\\
  J^2 &= - \sqrt{\frac{k+2}{2}} \partial \theta_2, \\
   \psi_1 \psi_3 &= \partial H,
\end{align}
and introduce a fourth free boson $X$ so to separate the $\vartheta_2 $
components
both in $\theta_2 $ and $H$:
\begin{align}
  \imath H &= \sqrt{ \frac{2}{k}} \vartheta_2 +  \imath \sqrt{\frac{k+2}{k}}
  X,
\\
  \theta_2 &= \sqrt{\frac{2}{k}} \left( \sqrt{\frac{k+2}{2}} \vartheta_2 +
\imath X\right).
\end{align}

A primary field $\Phi_{j \mu \tilde \mu }$ of the bosonic $SL
\left( 2, \setR \right)_{k+2}$ with eigenvalue $\mu $ with respect
to $J^2 $ and $\bar \mu $ with respect to $\bar J^2$ obeys by
definition
\begin{subequations}
  \begin{align}
    J^2 \left( z \right) \Phi_{j \mu \bar \mu } \left( w, \bar w  \right) &\sim
\frac{\mu
      \Phi_{j \mu \bar \mu } \left( w, \bar w  \right)}{z - w}, \\
    \bar J^2 \left( \bar z \right) \Phi_{j \mu \bar \mu } \left(  w, \bar w
\right)
    &\sim \frac{ \bar \mu
      \Phi_{j \mu \bar \mu } \left(  w, \bar w  \right)}{\bar z - \bar
      w}.
  \end{align}
\end{subequations}
Since $\Phi_{j \mu \bar \mu }$ is purely bosonic, the same
relation holds for the supercurrent:
\begin{equation}
  \mathcal{J}_2 \left( z \right) \Phi_{j \mu \bar \mu } \left( w , \bar w
\right) \sim \frac{\mu
    \Phi_{j \mu \bar \mu } \left( w, \bar w  \right)}{z - w}.
\end{equation}
Consider now the holomorphic part of $\Phi_{j \mu \bar \mu }
\left( z, \bar z \right)$. If $\Phi_{j \mu } $ is viewed as a
primary in the \textsc{swzw} model, we can use the parafermion
decomposition as follows:
\begin{equation}
  \label{eq:superparafermion}
  \Phi_{j\mu } \left( z \right) = U_{j \mu } \left( z \right) {\rm e}^{\imath \mu
\sqrt{2/k} \vartheta_2},
\end{equation}
where $U_{j \mu } \left( z \right)$ is a primary of the superconformal
$\nicefrac{SL \left( 2, \setR \right)_k}{U \left( 1 \right)}$. On the other
hand, we can just consider the bosonic \textsc{wzw} and write:
\begin{equation}
  \Phi_{j\mu } \left( z \right) = V_{j \mu } \left( z \right) {\rm e}^{\imath \mu
\sqrt{2/
      \left( k+2\right)} \theta_2} = V_{j \mu} \left( z \right){\rm e}^{\imath
\frac{2m}{k+2}
    \sqrt{\frac{k+2}{k}} X + \imath \mu \sqrt{2/k} \vartheta_2},
\end{equation}
where now $V_{j \mu } \left( z \right)$ is a primary of the bosonic
$\nicefrac{SL \left( 2, \setR\right)_{k+2}}{U \left( 1 \right)}$. The
scaling dimension for this latter operator (\emph{i.e.} its eigenvalue with
respect to $L_0$) is then given by:
\begin{equation}
\label{eq:boson-coset}
  \Delta \left( V_{j \mu} \right) = - \frac{j \left( j+1\right)}{k} -
\frac{\mu^2}{k+2}.
\end{equation}
An operator in the full supersymmetric $SL \left( 2, \setR \right)_k$ theory
is then obtained by adding the $\psi^1 \psi^3$ fermionic superpartner
contribution:
\begin{equation}
  \Phi_{j \mu \nu } \left( z \right) = \Phi_{j \mu } \left( z\right) {\rm e}^{\imath
\nu H} =
  V_{j \mu } \left( z \right) {\rm e}^{\imath \left( \frac{2 \mu }{k+2} + \nu \right)
    \sqrt{\frac{k+2}{k}} X } {\rm e}^{\imath
    \sqrt{2/k} \left( \mu
      +  \nu \right) \vartheta_2}
\end{equation}
that is an eigenvector of $\mathcal{J}_2 $ with eigenvalue $\mu +\nu $ where
$\mu \in \setR$ and $\nu $ can be decomposed as $\nu = n + a/2$ with $n \in
\setN$ and $a \in \setZ_2 $ depending on whether we consider the \textsc{ns}
or \textsc{r} sector. The resulting spectrum can be read directly as:
\begin{multline}
\label{eq:left-weights}
  \Delta \left( \Phi_{j \mu n} \left( z \right)\right) = - \frac{j \left(
      j+1\right)}{k} - \frac{\mu^2}{k+2} - \frac{k+2}{2k} \left( \frac{2\mu
    }{k+2} + n+\frac{a}{2} \right)^2 + \frac{1}{k} \left( \mu + n
    + \frac{a}{2}\right)^2 = \\
 = - \frac{j \left( j+1\right)}{k} - \frac{1}{2} \left( n +
    \frac{a}{2}\right)^2.
\end{multline}
Of course the last expression was to be expected since it is the sum of the
$\mathfrak{sl} \left( 2, \setR \right)_{k+2}$ Casimir and the contribution
of a light-cone fermion. Nevertheless the preceding construcion is useful
since it allowed us to isolate the $\mathcal{J}_2 $ contribution to the
spectrum $\left( \mu + \nu \right)^2/k$.

The right-moving part of the spectrum is somewhat simpler since there are no
superpartners. This means that we can repeat our construction above and the
eigenvalue of the $\bar L_0 $ operator is simply obtained by adding to the
dimension in Eq.~\eqref{eq:boson-coset} the contribution of the $\bar J^2 $
operator and of some $U \left( 1 \right)$ coming from the gauge sector:
\begin{equation}
\label{eq:right-weights}
  \bar \Delta \left( \bar \Phi_{j \bar \mu \bar n } \left( \bar z\right) \right)
=
  - \frac{j \left(
      j+1\right)}{k} - \frac{\bar \mu^2}{ k + 2 } + \left\{\frac{\bar \mu^2}{ k
+ 2 }
  + \frac{1}{k_g} \left( \bar n + \frac{\bar a}{2}\right)^2
  \right\},
\end{equation}
where again $\bar n \in \setN$ and $\bar a \in \setZ_2$ depending on the
sector.


\bibliography{Biblia}

\providecommand{\href}[2]{#2}\begingroup\raggedright\begin{thebibliography}{10}

\bibitem{Antoniadis:1990mn}
I.~Antoniadis, C.~Bachas, and A.~Sagnotti, {\it Gauged supergravity vacua in
  string theory},  {\em Phys. Lett.} {\bf B235} (1990) 255.

\bibitem{Petropoulos:1990fc}
P.~M.~S. Petropoulos, {\it Comments on su(1,1) string theory},  {\em Phys.
  Lett.} {\bf B236} (1990) 151.

\bibitem{Boonstra:1998yu}
H.~J. Boonstra, B.~Peeters, and K.~Skenderis, {\it Brane intersections, anti-de
  {S}itter spacetimes and dual superconformal theories},  {\em Nucl. Phys.}
  {\bf B533} (1998) 127--162,
  [\href{http://xxx.lanl.gov/abs/hep-th/9803231}{{\tt hep-th/9803231}}].

\bibitem{Maldacena:1998bw}
J.~M. Maldacena and A.~Strominger, {\it Ads(3) black holes and a stringy
  exclusion principle},  {\em JHEP} {\bf 12} (1998) 005,
  [\href{http://xxx.lanl.gov/abs/hep-th/9804085}{{\tt hep-th/9804085}}].

\bibitem{Giveon:1998ns}
A.~Giveon, D.~Kutasov, and N.~Seiberg, {\it Comments on string theory on
  ads(3)},  {\em Adv. Theor. Math. Phys.} {\bf 2} (1998) 733--780,
  [\href{http://xxx.lanl.gov/abs/hep-th/9806194}{{\tt hep-th/9806194}}].

\bibitem{Israel:2003ry}
D.~Isra{\"e}l, C.~Kounnas, and M.~P. Petropoulos, {\it Superstrings on ns5
  backgrounds, deformed ads(3) and holography},  {\em JHEP} {\bf 10} (2003)
  028, [\href{http://xxx.lanl.gov/abs/hep-th/0306053}{{\tt hep-th/0306053}}].

\bibitem{Banados:1992wn}
M.~Banados, C.~Teitelboim, and J.~Zanelli, {\it The black hole in
  three-dimensional space-time},  {\em Phys. Rev. Lett.} {\bf 69} (1992)
  1849--1851, [\href{http://xxx.lanl.gov/abs/hep-th/9204099}{{\tt
  hep-th/9204099}}].

\bibitem{Banados:1993gq}
M.~Banados, M.~Henneaux, C.~Teitelboim, and J.~Zanelli, {\it Geometry of the
  (2+1) black hole},  {\em Phys. Rev.} {\bf D48} (1993) 1506--1525,
  [\href{http://xxx.lanl.gov/abs/gr-qc/9302012}{{\tt gr-qc/9302012}}].

\bibitem{Bieliavsky:2002ki}
P.~Bieliavsky, M.~Rooman, and P.~Spindel, {\it Regular poisson structures on
  massive non-rotating btz black holes},  {\em Nucl. Phys.} {\bf B645} (2002)
  349--364, [\href{http://xxx.lanl.gov/abs/hep-th/0206189}{{\tt
  hep-th/0206189}}].

\bibitem{Bieliavsky:2003de}
P.~Bieliavsky, S.~Detournay, M.~Herquet, M.~Rooman, and P.~Spindel, {\it Global
  geometry of the 2+1 rotating black hole},  {\em Phys. Lett.} {\bf B570}
  (2003) 231--240, [\href{http://xxx.lanl.gov/abs/hep-th/0306293}{{\tt
  hep-th/0306293}}].

\bibitem{Chaudhuri:1989qb}
S.~Chaudhuri and J.~A. Schwartz, {\it A criterion for integrably marginal
  operators},  {\em Phys. Lett.} {\bf B219} (1989) 291.

\bibitem{Hassan:1992gi}
S.~F. Hassan and A.~Sen, {\it Marginal deformations of wznw and coset models
  from o(d,d) transformation},  {\em Nucl. Phys.} {\bf B405} (1993) 143--165,
  [\href{http://xxx.lanl.gov/abs/hep-th/9210121}{{\tt hep-th/9210121}}].

\bibitem{Giveon:1994ph}
A.~Giveon and E.~Kiritsis, {\it Axial vector duality as a gauge symmetry and
  topology change in string theory},  {\em Nucl. Phys.} {\bf B411} (1994)
  487--508, [\href{http://xxx.lanl.gov/abs/hep-th/9303016}{{\tt
  hep-th/9303016}}].

\bibitem{Forste:2003km}
S.~F\"orste and D.~Roggenkamp, {\it Current current deformations of conformal
  field theories, and wzw models},  {\em JHEP} {\bf 05} (2003) 071,
  [\href{http://xxx.lanl.gov/abs/hep-th/0304234}{{\tt hep-th/0304234}}].

\bibitem{Israel:2003cx}
D.~Isra{\"e}l, {\it Quantization of heterotic strings in a {G\"odel/anti de
  Sitter} spacetime and chronology protection},  {\em JHEP} {\bf 01} (2004)
  042, [\href{http://xxx.lanl.gov/abs/hep-th/0310158}{{\tt hep-th/0310158}}].

\bibitem{Israel:2004vv}
D.~Isra{\"e}l, C.~Kounnas, D.~Orlando, and P.~M. Petropoulos, {\it Electric /
  magnetic deformations of s**3 and ads(3), and geometric cosets},  {\em
  Fortsch. Phys.} {\bf 53} (2005) 73--104,
  [\href{http://xxx.lanl.gov/abs/hep-th/0405213}{{\tt hep-th/0405213}}].

\bibitem{Israel:2004cd}
D.~Israel, C.~Kounnas, D.~Orlando, and P.~M. Petropoulos, {\it Heterotic
  strings on homogeneous spaces},
  \href{http://xxx.lanl.gov/abs/hep-th/0412220}{{\tt hep-th/0412220}}.

\bibitem{Horne:1991gn}
J.~H. Horne and G.~T. Horowitz, {\it Exact black string solutions in
  three-dimensions},  {\em Nucl. Phys.} {\bf B368} (1992) 444--462,
  [\href{http://xxx.lanl.gov/abs/hep-th/9108001}{{\tt hep-th/9108001}}].

\bibitem{Knizhnik:1984nr}
V.~G. Knizhnik and A.~B. Zamolodchikov, {\it Current algebra and wess-zumino
  model in two dimensions},  {\em Nucl. Phys.} {\bf B247} (1984) 83--103.

\bibitem{Leutwyler:1991tv}
H.~Leutwyler and M.~A. Shifman, {\it Perturbation theory in the
  wess-zumino-novikov-witten model},  {\em Int. J. Mod. Phys.} {\bf A7} (1992)
  795--842.

\bibitem{Tseytlin:1992ri}
A.~A. Tseytlin, {\it Effective action of gauged wzw model and exact string
  solutions},  {\em Nucl. Phys.} {\bf B399} (1993) 601--622,
  [\href{http://xxx.lanl.gov/abs/hep-th/9301015}{{\tt hep-th/9301015}}].

\bibitem{Gepner:1986hr}
D.~Gepner and Z.-a. Qiu, {\it Modular invariant partition functions for
  parafermionic field theories},  {\em Nucl. Phys.} {\bf B285} (1987) 423.

\bibitem{Gepner:1987sm}
D.~Gepner, {\it New conformal field theories associated with lie algebras and
  their partition functions},  {\em Nucl. Phys.} {\bf B290} (1987) 10.

\bibitem{Tseytlin:1994my}
A.~A. Tseytlin, {\it Conformal sigma models corresponding to gauged
  {Wess-Zumino- Witten} theories},  {\em Nucl. Phys.} {\bf B411} (1994)
  509--558, [\href{http://xxx.lanl.gov/abs/hep-th/9302083}{{\tt
  hep-th/9302083}}].

\bibitem{Forste:1994wp}
S.~F\"orste, {\it A truly marginal deformation of sl(2, r) in a null
  direction},  {\em Phys. Lett.} {\bf B338} (1994) 36--39,
  [\href{http://xxx.lanl.gov/abs/hep-th/9407198}{{\tt hep-th/9407198}}].

\bibitem{Kiritsis:2003cx}
E.~Kiritsis, C.~Kounnas, P.~M. Petropoulos, and J.~Rizos, {\it Five-brane
  configurations, conformal field theories and the strong-coupling problem},
  \href{http://xxx.lanl.gov/abs/hep-th/0312300}{{\tt hep-th/0312300}}.

\bibitem{Horowitz:1995rf}
G.~T. Horowitz and A.~A. Tseytlin, {\it A new class of exact solutions in
  string theory},  {\em Phys. Rev.} {\bf D51} (1995) 2896--2917,
  [\href{http://xxx.lanl.gov/abs/hep-th/9409021}{{\tt hep-th/9409021}}].

\bibitem{Kiritsis:1994ta}
E.~Kiritsis and C.~Kounnas, {\it Infrared regularization of superstring theory
  and the one loop calculation of coupling constants},  {\em Nucl. Phys.} {\bf
  B442} (1995) 472--493, [\href{http://xxx.lanl.gov/abs/hep-th/9501020}{{\tt
  hep-th/9501020}}].

\bibitem{Rooman:1998xf}
M.~Rooman and P.~Spindel, {\it Goedel metric as a squashed anti-de sitter
  geometry},  {\em Class. Quant. Grav.} {\bf 15} (1998) 3241--3249,
  [\href{http://xxx.lanl.gov/abs/gr-qc/9804027}{{\tt gr-qc/9804027}}].

\bibitem{Witten:1991yr}
E.~Witten, {\it On string theory and black holes},  {\em Phys. Rev.} {\bf D44}
  (1991) 314--324.

\bibitem{Dijkgraaf:1992ba}
R.~Dijkgraaf, H.~Verlinde, and E.~Verlinde, {\it String propagation in a black
  hole geometry},  {\em Nucl. Phys.} {\bf B371} (1992) 269--314.

\bibitem{Gershon:1991qp}
D.~Gershon, {\it Exact solutions of four-dimensional black holes in string
  theory},  {\em Phys. Rev.} {\bf D51} (1995) 4387--4393,
  [\href{http://xxx.lanl.gov/abs/hep-th/9202005}{{\tt hep-th/9202005}}].

\bibitem{Horava:1991am}
P.~Horava, {\it Some exact solutions of string theory in four-dimensions and
  five-dimensions},  {\em Phys. Lett.} {\bf B278} (1992) 101--110,
  [\href{http://xxx.lanl.gov/abs/hep-th/9110067}{{\tt hep-th/9110067}}].

\bibitem{Klimcik:1994wp}
C.~Klimcik and A.~A. Tseytlin, {\it Exact four-dimensional string solutions and
  toda like sigma models from 'null gauged' wznw theories},  {\em Nucl. Phys.}
  {\bf B424} (1994) 71--96, [\href{http://xxx.lanl.gov/abs/hep-th/9402120}{{\tt
  hep-th/9402120}}].

\bibitem{Horne:1991cn}
J.~H. Horne, G.~T. Horowitz, and A.~R. Steif, {\it An equivalence between
  momentum and charge in string theory},  {\em Phys. Rev. Lett.} {\bf 68}
  (1992) 568--571, [\href{http://xxx.lanl.gov/abs/hep-th/9110065}{{\tt
  hep-th/9110065}}].

\bibitem{Horowitz:1993jc}
G.~T. Horowitz and D.~L. Welch, {\it Exact three-dimensional black holes in
  string theory},  {\em Phys. Rev. Lett.} {\bf 71} (1993) 328--331,
  [\href{http://xxx.lanl.gov/abs/hep-th/9302126}{{\tt hep-th/9302126}}].

\bibitem{Israel:1967wq}
W.~Israel, {\it Event horizons in static vacuum space-times},  {\em Phys. Rev.}
  {\bf 164} (1967) 1776--1779.

\bibitem{Heusler:1998ua}
M.~Heusler, {\it Stationary black holes: Uniqueness and beyond},  {\em Living
  Rev. Rel.} {\bf 1} (1998) 6.

\bibitem{Gibbons:2002av}
G.~W. Gibbons, D.~Ida, and T.~Shiromizu, {\it Uniqueness and non-uniqueness of
  static black holes in higher dimensions},  {\em Phys. Rev. Lett.} {\bf 89}
  (2002) 041101, [\href{http://xxx.lanl.gov/abs/hep-th/0206049}{{\tt
  hep-th/0206049}}].

\bibitem{Gregory:1993vy}
R.~Gregory and R.~Laflamme, {\it Black strings and p-branes are unstable},
  {\em Phys. Rev. Lett.} {\bf 70} (1993) 2837--2840,
  [\href{http://xxx.lanl.gov/abs/hep-th/9301052}{{\tt hep-th/9301052}}].

\bibitem{Sfetsos:1992yi}
K.~Sfetsos, {\it Conformally exact results for sl(2,r) x so(1,1)(d-2) / so(1,1)
  coset models},  {\em Nucl. Phys.} {\bf B389} (1993) 424--444,
  [\href{http://xxx.lanl.gov/abs/hep-th/9206048}{{\tt hep-th/9206048}}].

\bibitem{Kiritsis:1995iu}
E.~Kiritsis and C.~Kounnas, {\it Infrared behavior of closed superstrings in
  strong magnetic and gravitational fields},  {\em Nucl. Phys.} {\bf B456}
  (1995) 699--731, [\href{http://xxx.lanl.gov/abs/hep-th/9508078}{{\tt
  hep-th/9508078}}].

\bibitem{Glenn:2001}
G.~Barnich and F.~Brandt, {\it Covariant theory of asymptotic symmetries,
  conservation laws and central charges},  {\em Nucl. Phys.} {\bf B633} (2002)
  3--82, [\href{http://xxx.lanl.gov/abs/hep-th/0111246}{{\tt hep-th/0111246}}].

\bibitem{Natsuume:1998ij}
M.~Natsuume and Y.~Satoh, {\it String theory on three dimensional black holes},
   {\em Int. J. Mod. Phys.} {\bf A13} (1998) 1229--1262,
  [\href{http://xxx.lanl.gov/abs/hep-th/9611041}{{\tt hep-th/9611041}}].

\bibitem{Hemming:2001we}
S.~Hemming and E.~Keski-Vakkuri, {\it The spectrum of strings on {BTZ} black
  holes and spectral flow in the {SL(2,R) WZW} model},  {\em Nucl. Phys.} {\bf
  B626} (2002) 363--376, [\href{http://xxx.lanl.gov/abs/hep-th/0110252}{{\tt
  hep-th/0110252}}].

\bibitem{Coussaert:1994tu}
O.~Coussaert and M.~Henneaux, {\it Self-dual solutions of 2+1 einstein gravity
  with a negative cosmological constant},
  \href{http://xxx.lanl.gov/abs/hep-th/9407181}{{\tt hep-th/9407181}}.

\bibitem{Giveon:2003wn}
A.~Giveon, A.~Konechny, A.~Pakman, and A.~Sever, {\it Type 0 strings in a 2-d
  black hole},  {\em JHEP} {\bf 10} (2003) 025,
  [\href{http://xxx.lanl.gov/abs/hep-th/0309056}{{\tt hep-th/0309056}}].

\end{thebibliography}\endgroup

\end{document}